\documentclass[]{aastex631}

\pdfoutput=1
\usepackage{amsmath}
\usepackage{amsfonts}
\usepackage{amssymb}
\usepackage{graphicx, rotating}
\usepackage{epsfig}
\usepackage{latexsym}
\usepackage{graphicx}
\usepackage{color}
\usepackage{amsmath,bm,amssymb}
\usepackage{ulem}

\usepackage{color}
\usepackage[english]{babel}
\usepackage{hyperref}

\renewcommand{\figurename}{Fig.}
\renewcommand{\tablename}{Table.}
\newcommand{\Slash}[1]{{\ooalign{\hfil#1\hfil\crcr\raise.167ex\hbox{/}}}}

\newcommand{\beq}{\begin{equation}}  \newcommand{\eeq}{\end{equation}}
\newcommand{\bef}{\begin{figure}}  \newcommand{\eef}{\end{figure}}
\newcommand{\bec}{\begin{center}}  \newcommand{\eec}{\end{center}}
  
\newcommand{\laq}[1]{\label{eq:#1}}

\def\({\left(}
\def\){\right)}

\def\O{\mathcal{O}}
\def\U{\mathop{\rm U}}

\newcommand{\AND}{~{\rm and}~}
\newcommand{\EV}{ {\rm \, eV} }
\newcommand{\KEV}{ {\rm \, keV} }

\newcommand{\GEV}{ {\rm \, GeV} }

\def\o{\over}
\def\a{\alpha}
\def\b{\beta}
\def\c{\varepsilon}
\def\d{\delta}
\def\e{\epsilon}
\def\f{\phi}
\def\g{\gamma}
\def\h{\theta}
\def\k{\kappa}
\def\l{\lambda}
\def\m{\mu}
\def\n{\nu}
\def\p{\psi}
\def\q{\partial}
\def\r{\rho}
\def\s{\sigma}
\def\t{\tau}
\def\u{\upsilon}
\def\w{\omega}
\def\x{\xi}
\def\y{\eta}
\def\z{\zeta}
\def\D{\Delta}
\def\G{\Gamma}
\def\F{\Phi}
\def\P{\Psi}
\def\S{\Sigma}

\def\tl{\tilde}
\def\*{\dagger}

\begin{document}

 \begin{flushright}
 TU-1191
 \end{flushright}

\title{Indirect Detection of Decaying Dark Matter with High Angular Resolution: 
\\Case for axion search by IRCS at Subaru Telescope}

\author[0000-0001-8785-6351]{Wen Yin}
\affiliation{Department of Physics, Tohoku University,\\ Sendai, Miyagi 980-8578, Japan }

\author[0000-0002-8758-8139]{Kohei Hayashi}
\affiliation{National Institute of Technology,\\ Sendai College, Natori, Miyagi 981-1239, Japan}
\affiliation{Astronomical Institute, Tohoku University,\\ Sendai, Miyagi 980-8578, Japan}
\affiliation{Institute for Cosmic Ray Research, The University of Tokyo,\\ Kashiwa, Chiba 277-8582, Japan}


\begin{abstract}

Recent advances in cosmic-ray detectors have provided exceptional sensitivities of dark matter with high angular resolution.
Motivated by this, we present a comprehensive study of cosmic-ray flux from dark matter decay in dwarf spheroidal galaxies (dSphs), with a focus on detectors possessing arcsecond-level field of view and/or angular resolution. We propose to use differential $D$-factors, which are estimated for various dSphs, since such detectors are sensitive to their dark matter distributions.  
 Our findings reveal that the resulting signal flux can experience a more than $\O$(1-10) enhancement with different theoretical uncertainty compared to traditional estimations. Based on this analysis, we find that the Infrared Camera and Spectrograph (IRCS) installed on the 8.2m Subaru telescope can be a  good  dark matter detector for the mass in the eV range, particularly axion-like particles (ALPs). Observing the Draco or Ursa Major II galaxies with the IRCS for just a few nights will be sufficient to surpass the stellar cooling bounds for ALP dark matter with a mass in the range of $1\EV \lesssim m_a \lesssim 2\EV$. 

\end{abstract}

% \keywords{Classical Novae (251) --- Ultraviolet astronomy(1736) --- History of astronomy(1868) --- Interdisciplinary astronomy(804)}

%% We recommend that authors also use the natbib \citepp
%% and \citept commands to identify citations.  The citations are

%%%%%%%%%%%%%%%%%%%%%%%%%%%%%%%%%%%%%%
\section{Introduction}

The success of the $\Lambda$ cold dark matter ($\Lambda$CDM) model strongly suggests that dark matter (DM) played an important role in the evolution of the early universe and the formation of galaxies. It is certainly present in the current universe, and, in addition, it is evidence beyond the Standard Model of elementary particles. However, their particle properties, such as mass and interactions (other than gravity), have remained unknown for about a century, despite various experimental search programs. DM is understood to be localized around the center of galaxies.  Taking advantage of this property, there have been two main types of search methods: direct detection experiments in which DM in Milky Way galaxy directly scatter with molecule leading to nucleon and electron recoils (XENONnT~\citep{XENON:2020kmp}, LZ~\citep{LZ:2022ufs}, PandaX~\citep{PANDA:2009yku} experiments, etc.), or there can be indirect detection experiments that observe DM induced cosmic rays in various galaxies (ATHENA\footnote{For more information on ATHENA, see https://www.the-athena-x-ray-observatory.eu. }, CTA~\citep{CTAConsortium:2017dvg}, Fermi-LAT~\citep{Fermi-LAT:2012edv,Fermi-LAT:2018lqt}, etc.) The former has excluded many possible DM candidates with masses above GeV, and the latter has eliminated many possible DM candidates above keV based on observations of X-rays and $\gamma$-rays. Recently many new experiments have been planned and performed in different mass ranges.

In particular, indirect detection started to be considered for lighter DM than keV by taking advantage of state-of-art detectors. 
For instance, the axion-like particle (ALP) DM candidate around the mass range of $2.7-5.3$eV is constrained from the real data of the MUSE optical spectrograph looking at the Leo T dwarf spheroidal galaxy~\citep{bacon2017muse,Regis:2020fhw} by requiring the suppressed DM flux compared to the actual data. In \citep{Grin:2006aw}, the authors use the VIMOS data looking at the galaxy clusters Abell 2667 and 2390 to constrain the DM  in the mass range of 4.5-7.7eV. They provide the strongest bound for the ALP DM around the mass range. Those detectors have a very good angular resolution of arcsec level. 

 More recently, it was pointed out that the state-of-art infrared spectrograph, such as the Warm INfrared Echelle spectrograph to Realize Extreme Dispersion and sensitivity \citep{ikeda2006winered, yasui2008warm, kondo2015warm, ikeda2016high, ikeda2018very, 2022WINERED} installed in the $6.5$m Magellan Clay telescope, and Near-Infrared Spectrograph in the James Webb Space Telescope\footnote{see  https://jwst.nasa.gov/index.html.} cannot only constrain the DM but also discover the DM by noting that the sky background can be highly suppressed due to the extremely good spectral resolution. 
This is the case that the DM in dSphs has the decay into  two particles involving a photon because then the resulting photon has an almost line-like spectrum while the background noise constructs continuous spectra that are darker in a detector with better energy resolution. 
It was also mentioned that the DM could be searched for in the Infrared Camera and Spectrograph (IRCS) at the 8.2m Subaru telescope~\citep{tokunaga1998infrared,kobayashi2000ircs}  with a similar method, and the good angular resolution of these detectors may have underestimated the DM signal from the denser galaxy center. However, a detailed study was not carried out.  These spectrographs can be very good DM detectors due to the nice energy resolution and angular resolution of $\O(0.01-0.1)\rm arcsec^2$.

In this paper, we conducted a detailed investigation of the indirect detection of DM with a detector with a small field of view or/and good angular resolution. 
In particular, we point out the importance of using the differential $D$-factor for the DM search in dSphs with those detectors and estimate the differential $D$-factor for various dSphs. It is important because such a detector is sensitive to the DM distribution. 
{In the estimation, we adopted the mass density profiles estimated in Refs.\,\citep{Hayashi:2020jze,Hayashi:2022wnw} where the authors used non-spherical mass models based on axisymmetric Jeans equations. In this case, one can alleviate the strong degeneracy between DM density and stellar velocity anisotropy to get a more robust prediction of the differential $D$-factor distribution around the galaxy centers.}
 By employing the differential $D$-factor, we find that the signal flux typically experiences a more than $\O(1-10)$ enhancement compared to the na\"{i}ve estimation and that the theoretical uncertainty is very different. 
By using our findings, we propose to search for eV mass range DM by using the IRCS at the Subaru telescope and show that the observation with a few nights is sufficient to exceed the stellar cooling bounds for ALP DM with mass $1\EV\lesssim m_a \lesssim 2\EV$.

\section{Differential $D$-factors in dwarf spheroidal galaxies}

Let us consider that the DM $\f$ decays with a decay width $\Gamma_\f$ into standard model particles that travel freely to the detector. 
The resulting differential flux of the signal cosmic ray, say the photon $\gamma$, from the decay, can be decomposed into two components:
\beq
\frac{\partial^2\F_\g}{\partial E_\g \partial\Omega}=  \frac{\partial^2\F_\g^{\text{extra}}}{\partial E_\g \partial\Omega}+\sum_i\frac{\partial^2\F_{\g,i}}{\partial E_\g \partial\Omega}
\eeq
The first term represents the extragalactic component which is isotropic and has a continuous spectrum. 
The second term represents the component from a nearby galaxy or galaxy cluster, $i$. 
It has a more localized spectrum, either in angular or in energy (due to the suppressed Doppler effect) than the extragalactic component.  In this paper, we focus on the second term, especially for the dSphs.

By neglecting the photon scattering or absorption during the propagation to Earth, the second component can be estimated from 
\beq
\frac{\partial^2 \Phi_{\g,i}}{\partial E_\g \partial  \Omega}=\int ds  \frac{1}{4\pi s^2}   \left(\frac{\Gamma_{\f}  \rho^{i}_\phi(s,  \Omega )}{m_\phi}\right) \, s^2\, \frac{\partial^2 N_{\f,i}}{\partial E_\gamma \partial \Omega }[ s, \Omega].
\eeq
Here, $s$ is the line of sight distance, $\Omega$ is the spherical coordinate, $\rho^{i}_\f$ and $\frac{\partial^2 N_{\f,i} }{\partial E_\g \partial \Omega} [s, \Omega]$ represent the DM energy density distribution and photon spectrum from a single DM decay around galaxy $i$, respectively, and both depend on the DM halo model and the properties of galaxy $i$.  

The photon spectrum depends on the detail of the DM model and the DM velocity distribution,
\beq
\frac{\partial^2 N_{\f,i} }{\partial E_\g \partial \Omega} [s, \Omega] =  \int{d^3\vec{v} f_i(\vec{v}, s ,\Omega ) F_{\rm rest}\(\frac{E_\g}{1-\vec{v} \cdot \vec{\Omega}}\)}
\eeq
Here, $f_i$ is the DM velocity distribution. $F_{\rm rest}[E]$ represents the number distribution of cosmic-ray from the decay of a DM particle at rest. $F_{\rm rest}$, itself, does not depend on the angular direction due to the rotational invariance.\footnote{We have implicitly assumed that the spin of the DM does not have a special direction in average and did not include the distribution of spin in $f_i$. In more general, if there is a dark magnetic field, the DM spin may be polarized. In this case, the resulting cosmic-ray flux from the decay can have more non-trivial angular dependence. }
For instance, if we consider a decay into two photons, we have $F_{\rm rest}[E_\gamma ]=2\delta(E_\gamma-m_\f/2 )$, with $m_\f$ being the mass of the DM. For more than two body decay, we have a broader distribution for $F_{\rm rest}[E].$ 
We have included the Doppler shift, which appears in the denominator of the argument of $F_{\rm rest}$. Together with $f_i$, the 
resulting $\frac{\partial^2 N_{\f,i} }{\partial E_\g \partial \Omega}$ is distorted from $F_{\rm rest}$.
In the dSphs, the velocity dispersion in $ f_i(\vec{v}, s,\Omega )$ is less than $10$km/s, and the distortion of the spectrum induced by the Doppler shift is smaller than $<10^{-4}\times m_\f/2.$ 
If we concentrate on the center of the dSph, which is our focus, the Doppler shift distortion on the spectrum is even suppressed (see, e.g., Fig.3 of Ref.\,\citep{Bessho:2022yyu}.) Since various state-of-art detectors have an energy resolution of $R=\O(10^{4})$,  
we neglect this distortion effect for simplicity. As a result, we get
\beq
\laq{simp}
\frac{\partial^2 N_{\f,i=\rm dSph} }{\partial E_\g \partial \Omega} \simeq F_{\rm rest}[\frac{E_\gamma}{1-v_i^{\rm rel}}]
\eeq
where $v_i^{\rm rel}$ is the radial velocity of the dSph $i$. This does not depend on $s$ and angular distance from the galactic center.

\begin{figure}
\begin{center}  
\includegraphics[width=75mm]{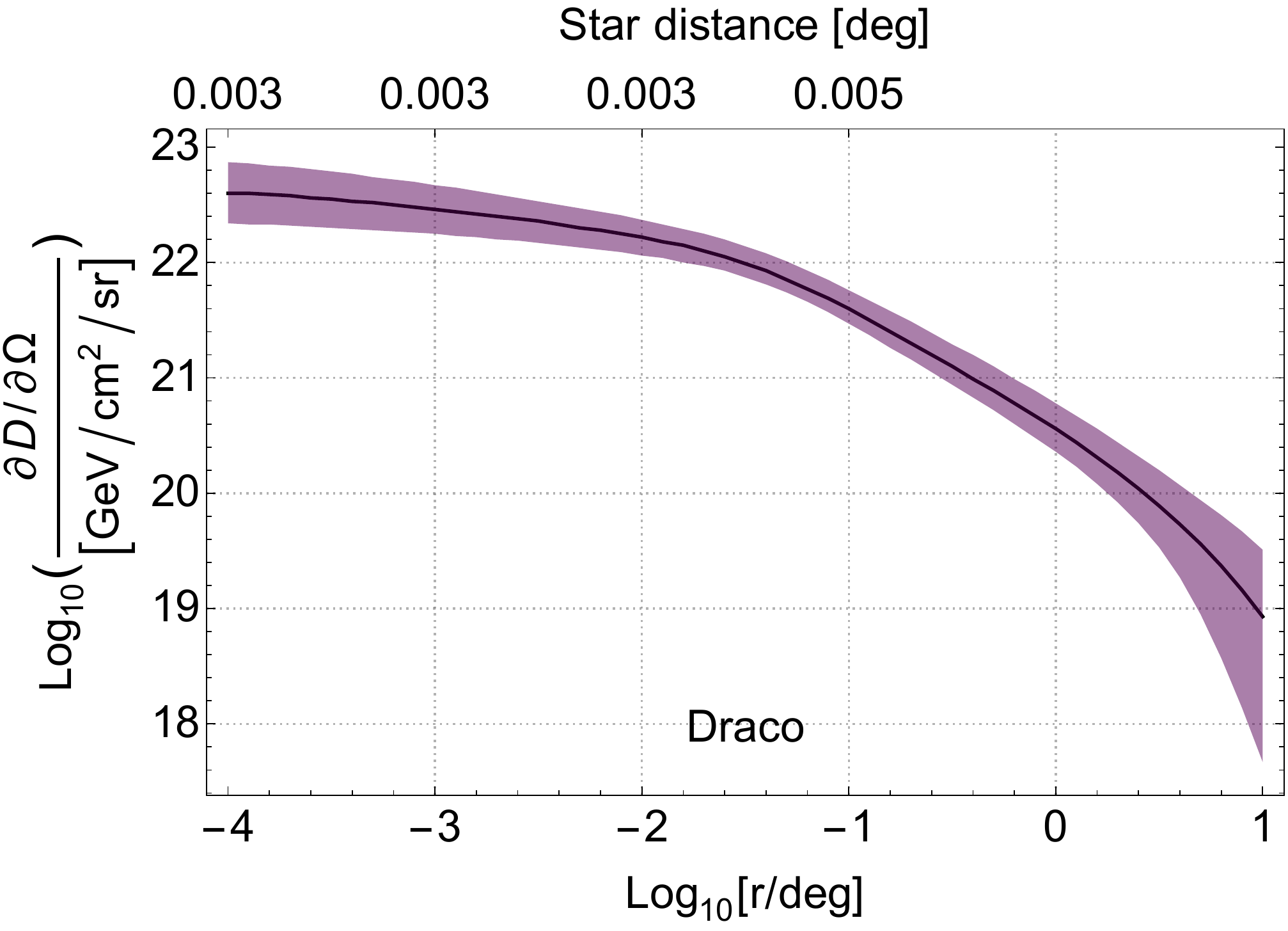}
\includegraphics[width=75mm]{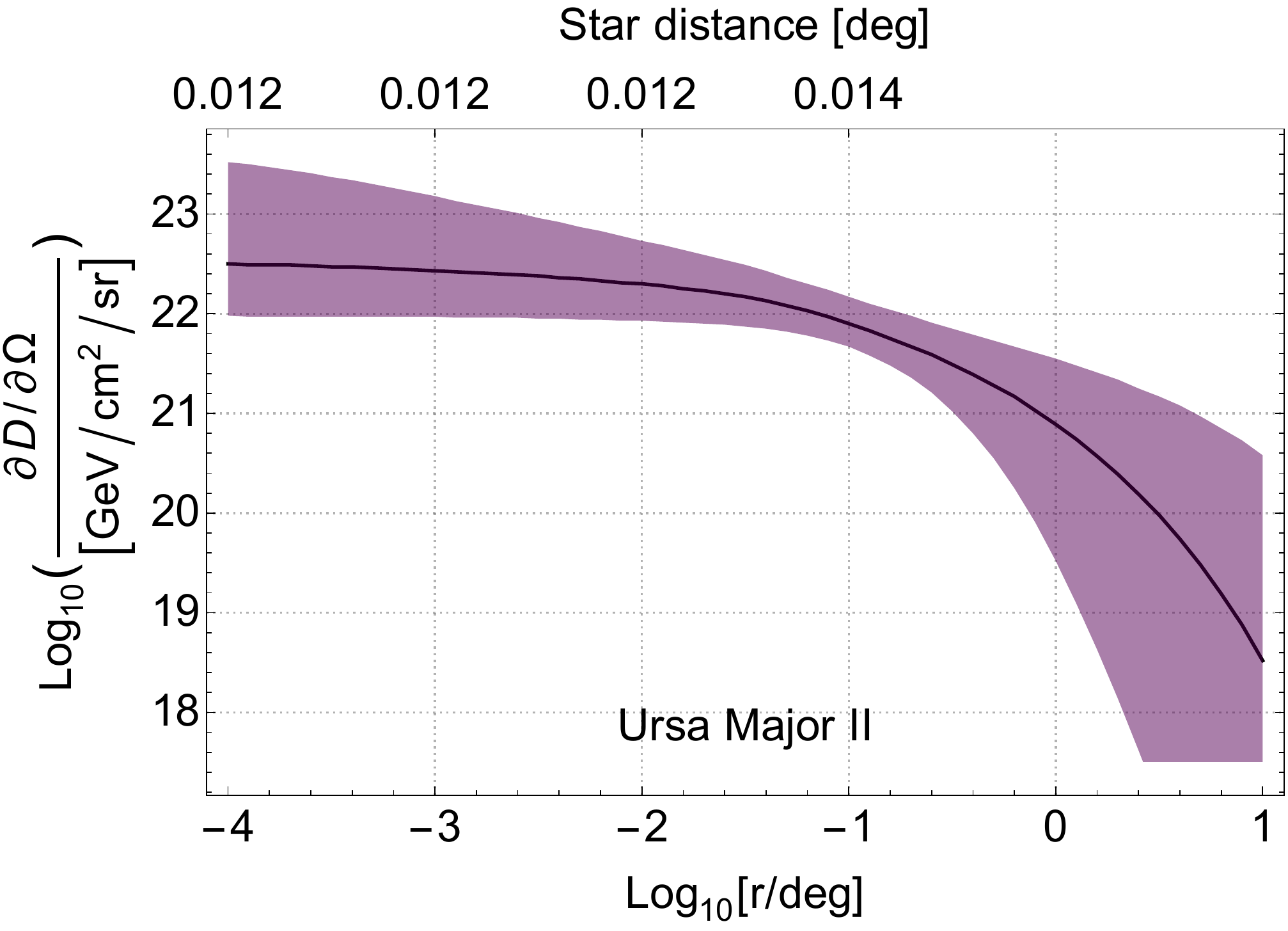}
\end{center}
\caption{The differential $D$-factor for Draco and Ursa Major II by varying the angular distance, $\log_{10}{(r /[\rm deg])}$, from the galactic center. 
We also show the typical angular distance, $\bar{d}_{\rm stars}$, between two visible stars in [deg] in the top frame.
   } \label{fig:1}
\end{figure}

\begin{figure}[!t]
\begin{center}  
\includegraphics[width=45mm]{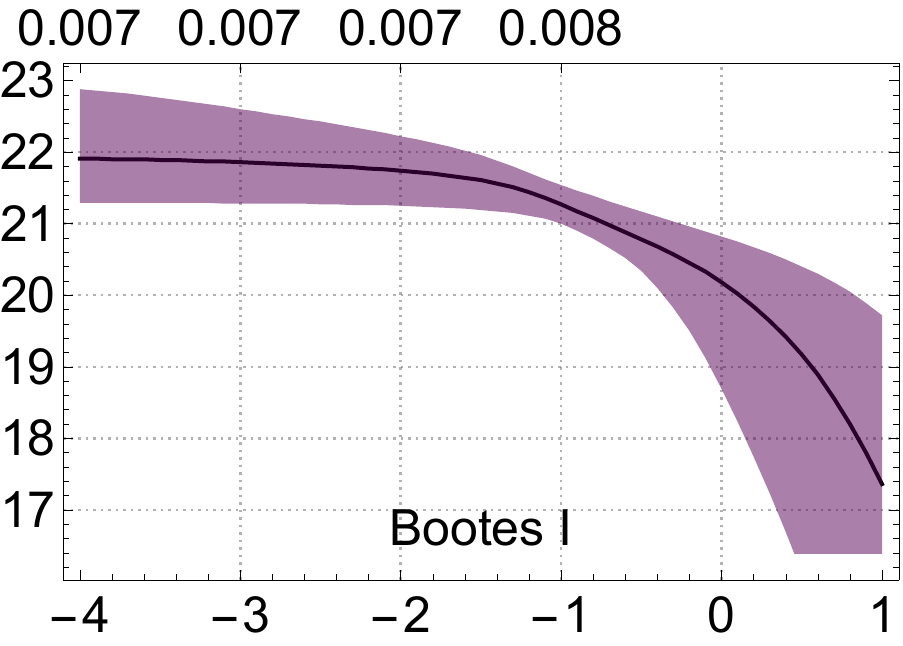}
\includegraphics[width=45mm]{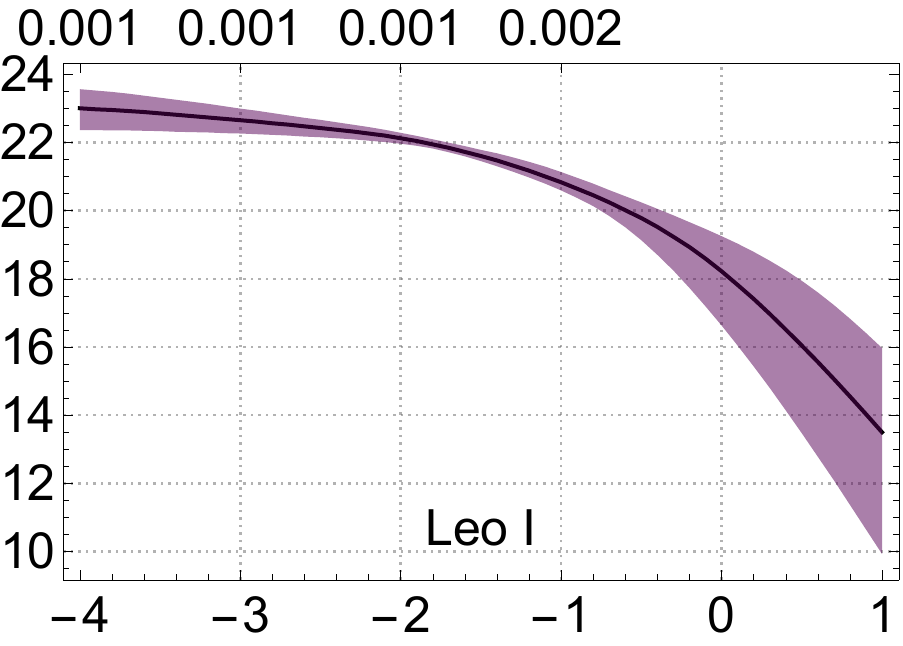}
\includegraphics[width=45mm]{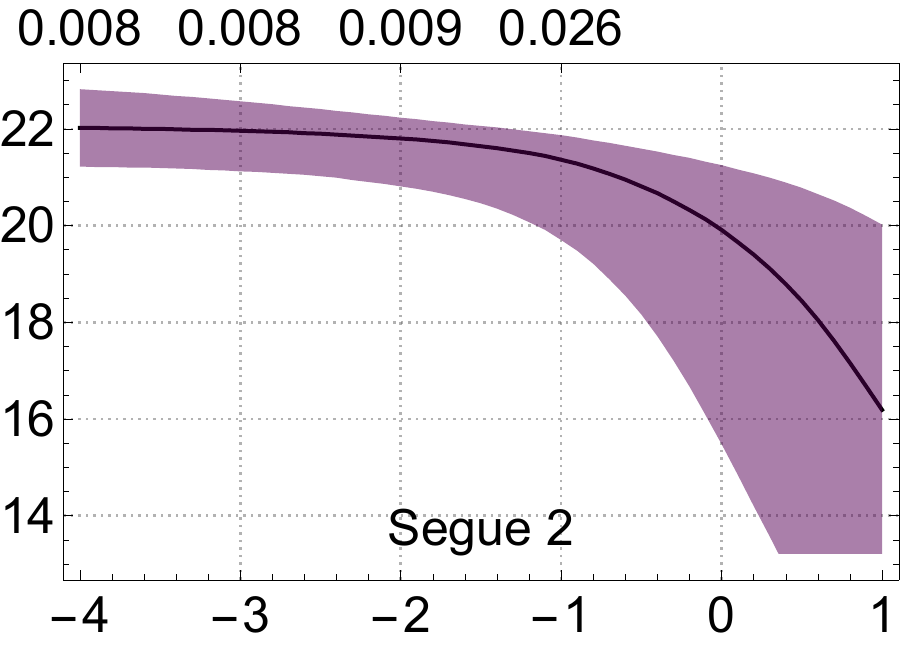}
\includegraphics[width=45mm]{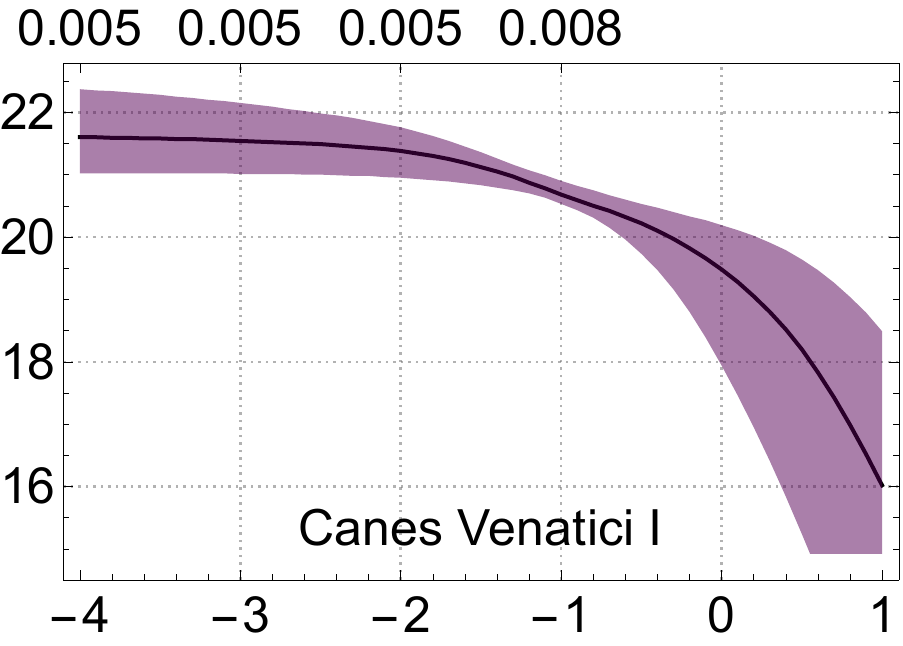}
\includegraphics[width=45mm]{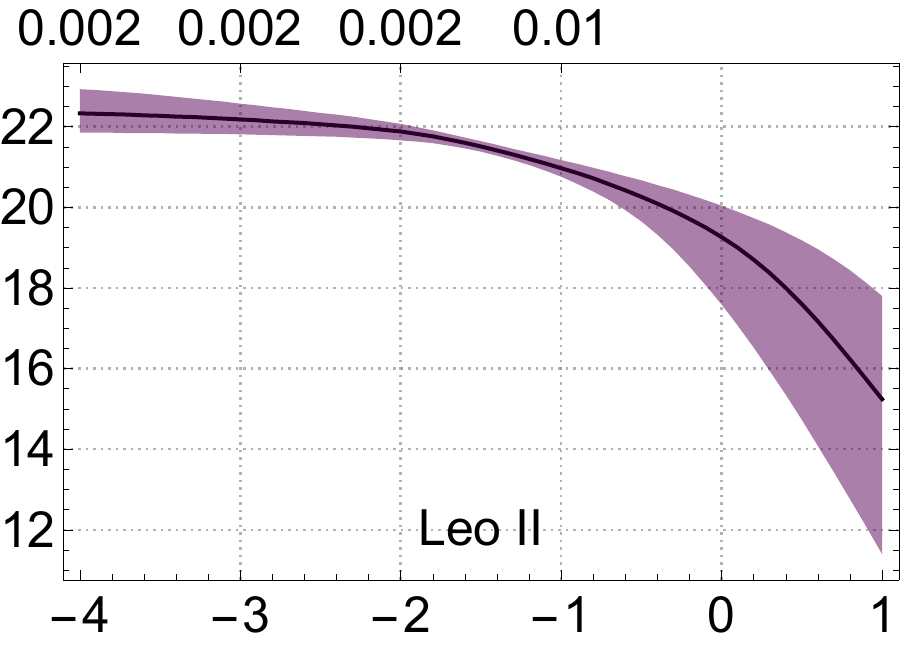}
\includegraphics[width=45mm]{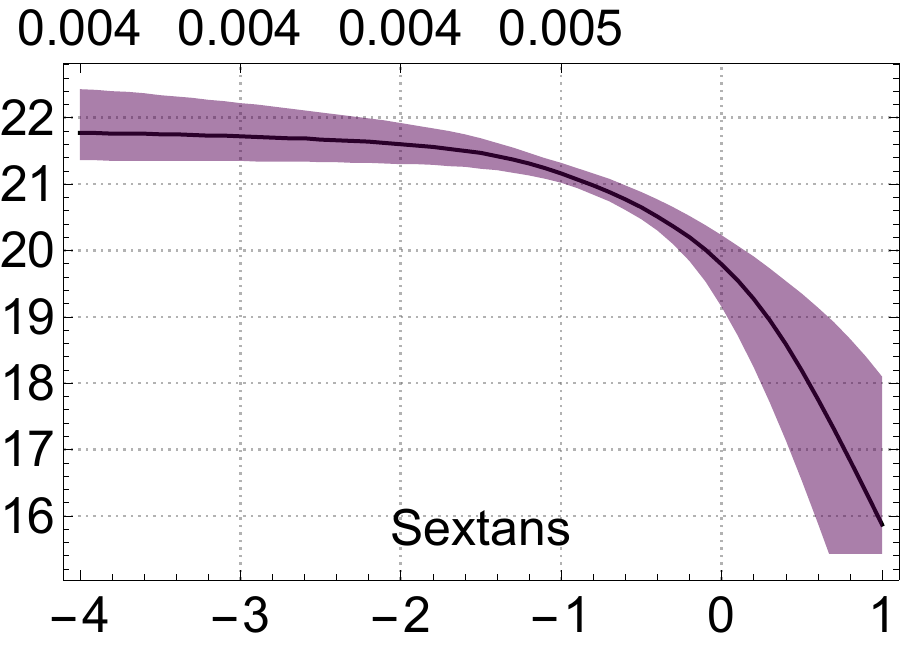}
\includegraphics[width=45mm]{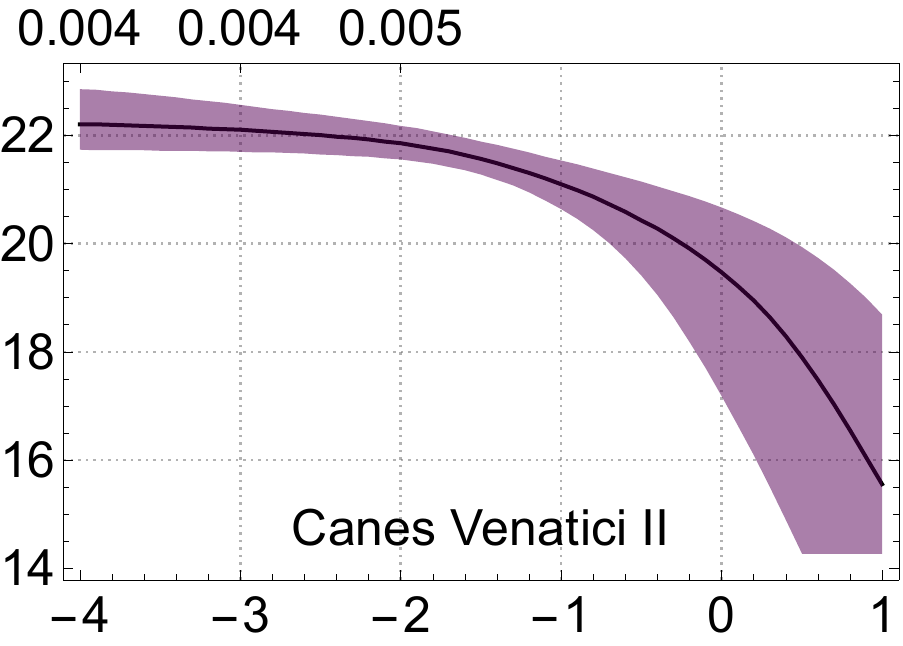}
\includegraphics[width=45mm]{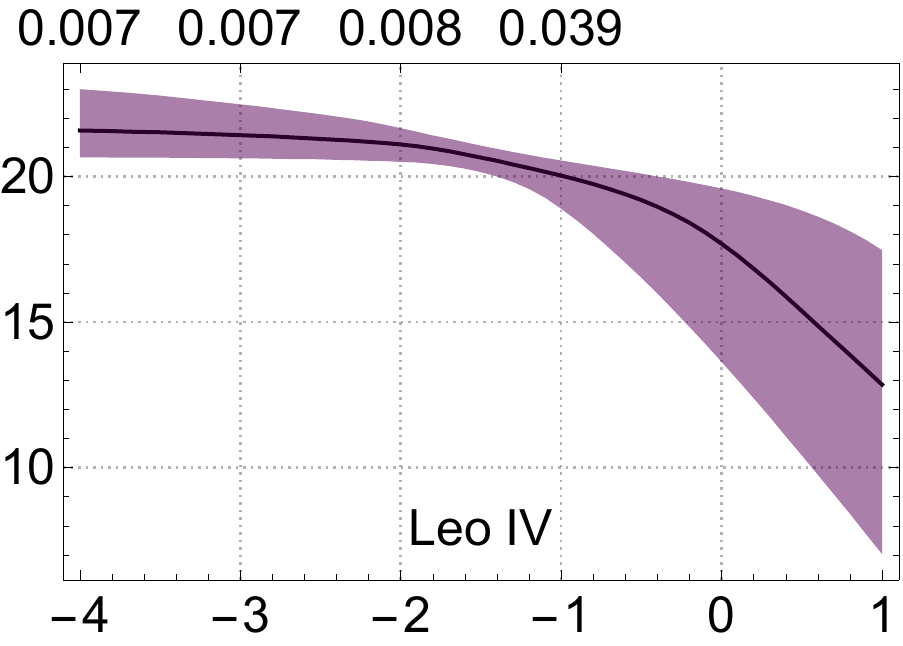}
\includegraphics[width=45mm]{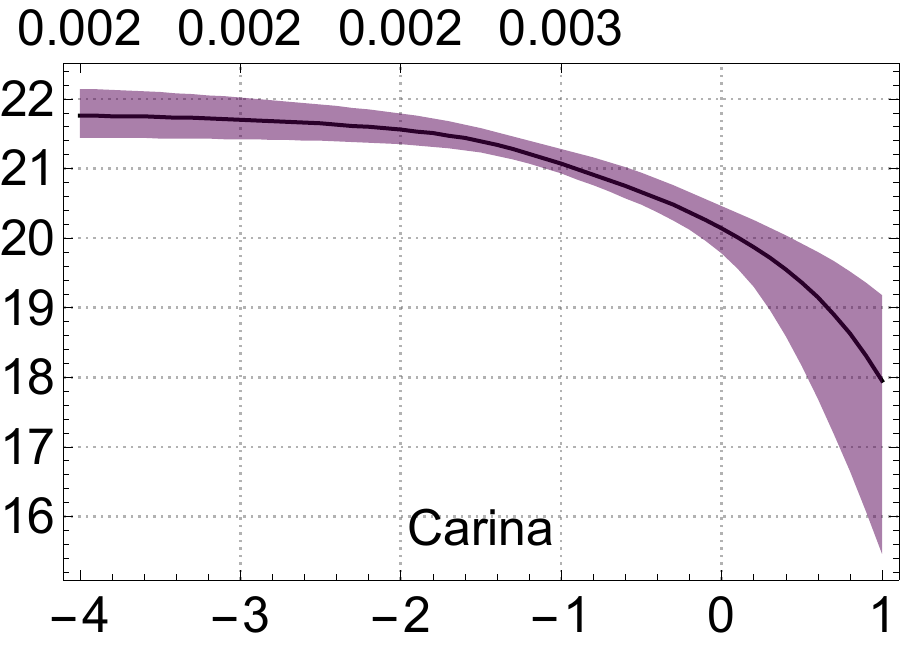}
\includegraphics[width=45mm]{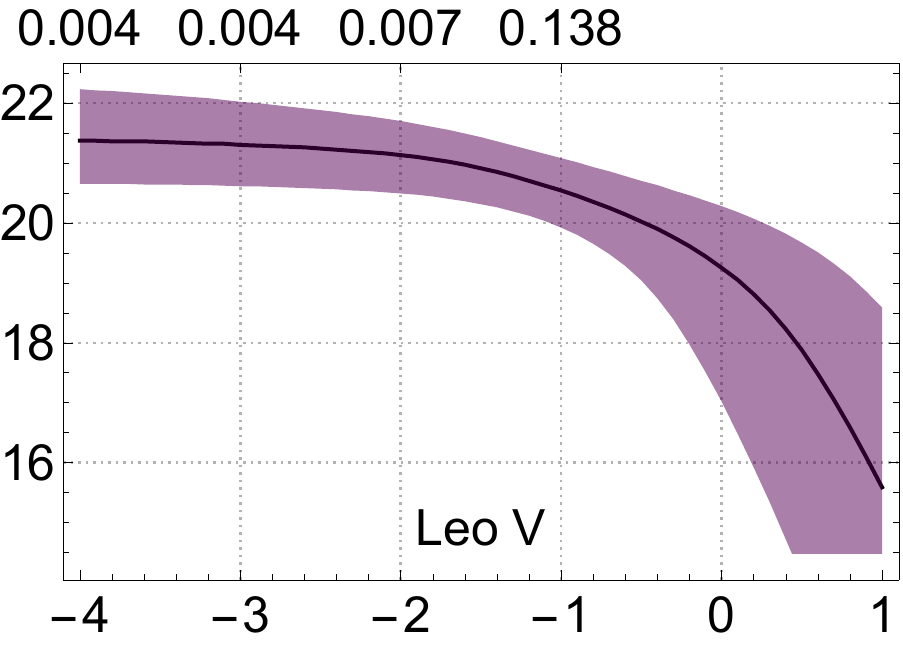}
\end{center}
\caption{Same as Fig.\ref{fig:1} but for different dSphs. For the labels of the axes see Fig.\ref{fig:1}.} \label{fig:2}
\end{figure}

\begin{figure}[!t]
\begin{center}  
\includegraphics[width=45mm]{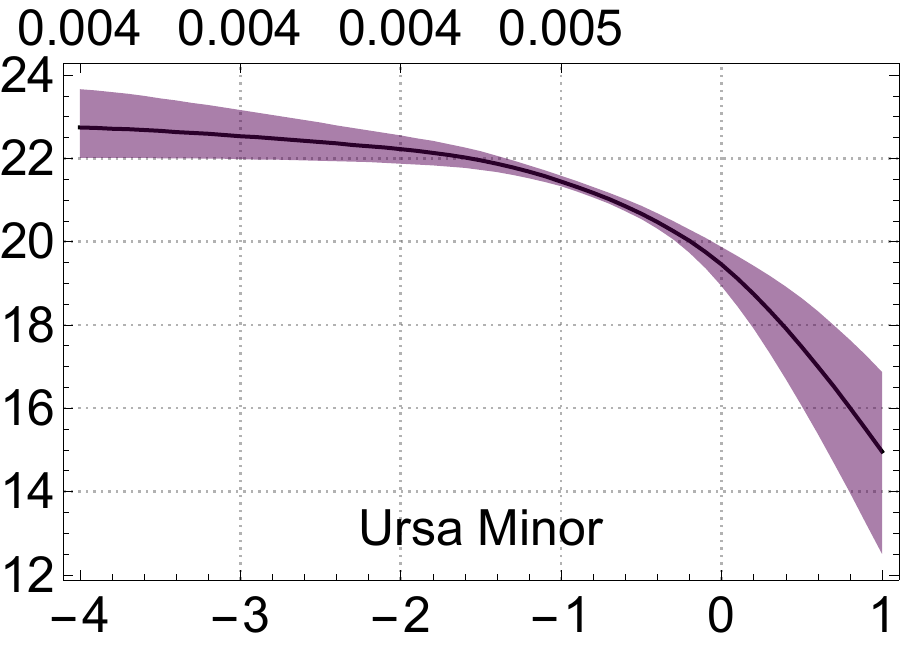}
\includegraphics[width=45mm]{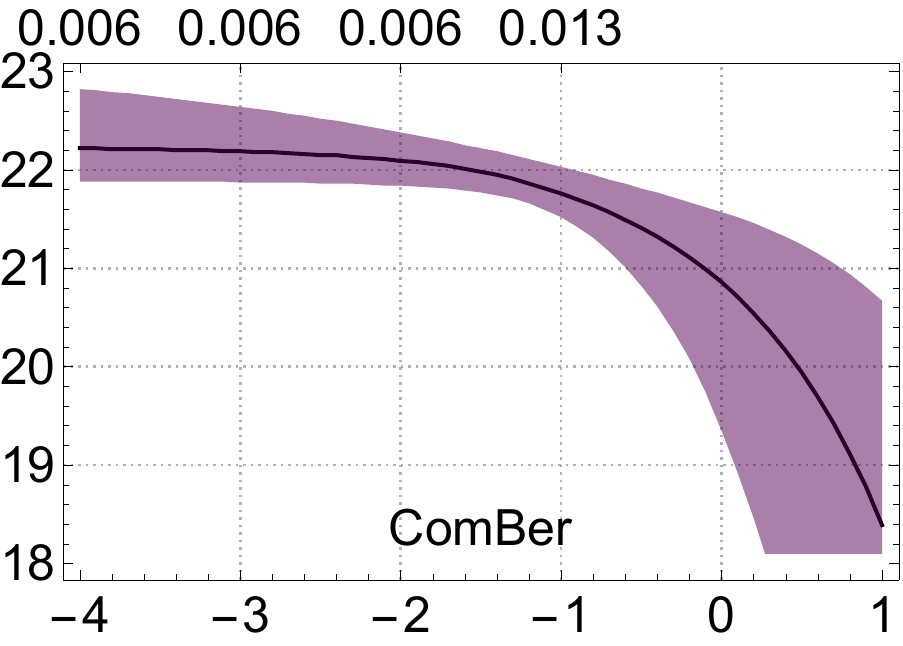}
\includegraphics[width=45mm]{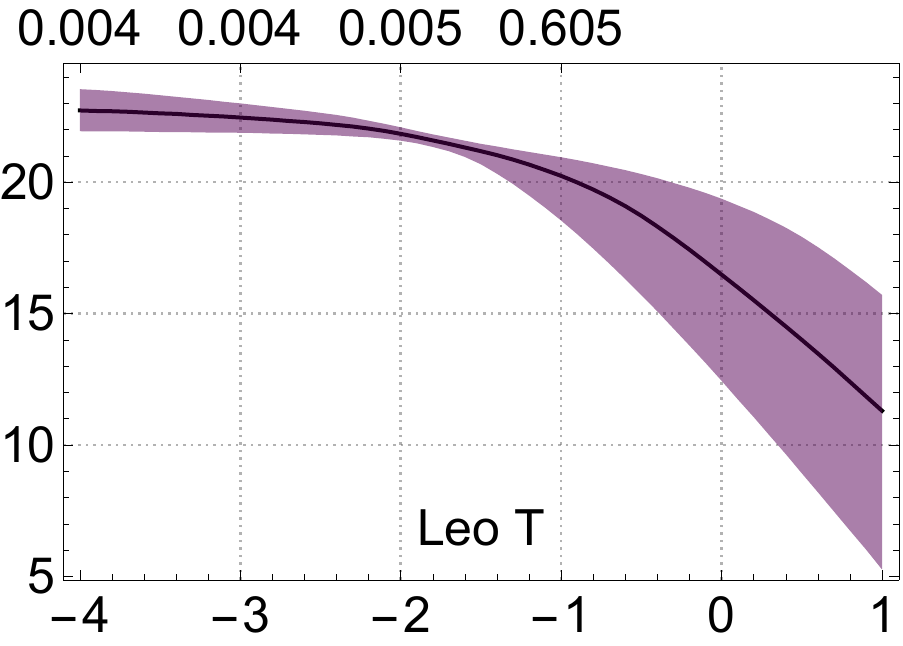}
\includegraphics[width=45mm]{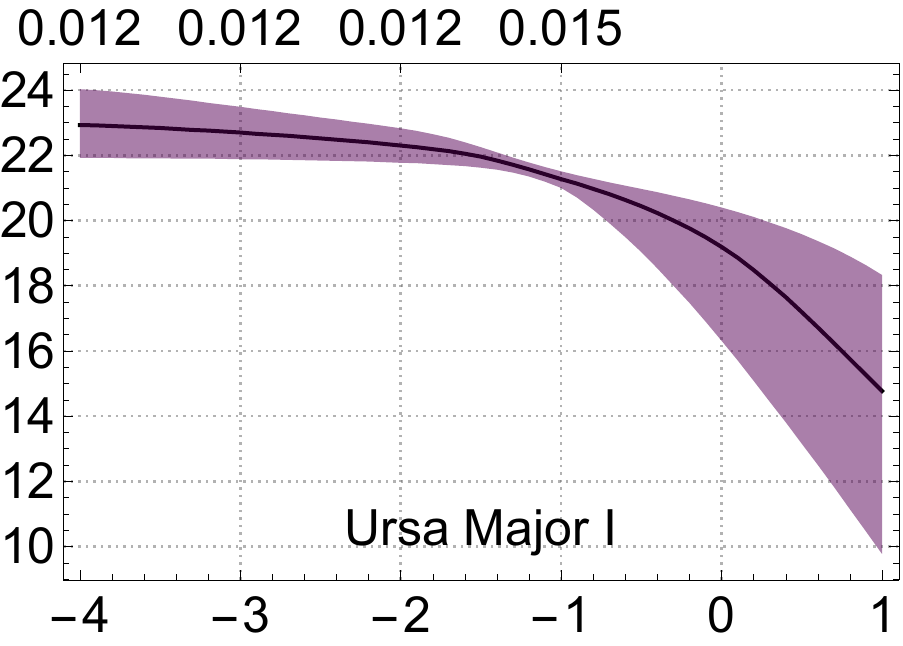}
\includegraphics[width=45mm]{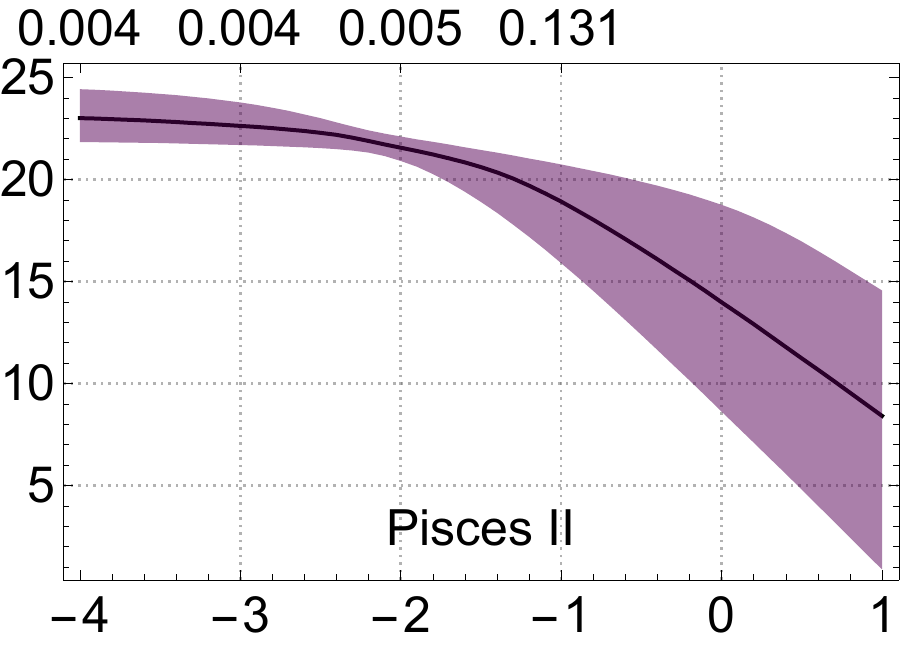}
\includegraphics[width=45mm]{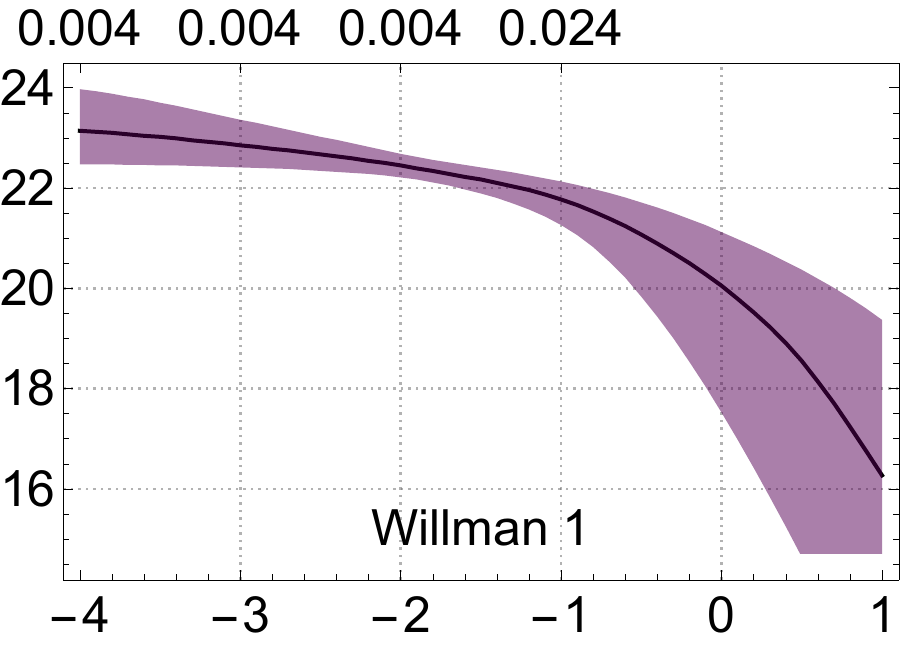}
\includegraphics[width=45mm]{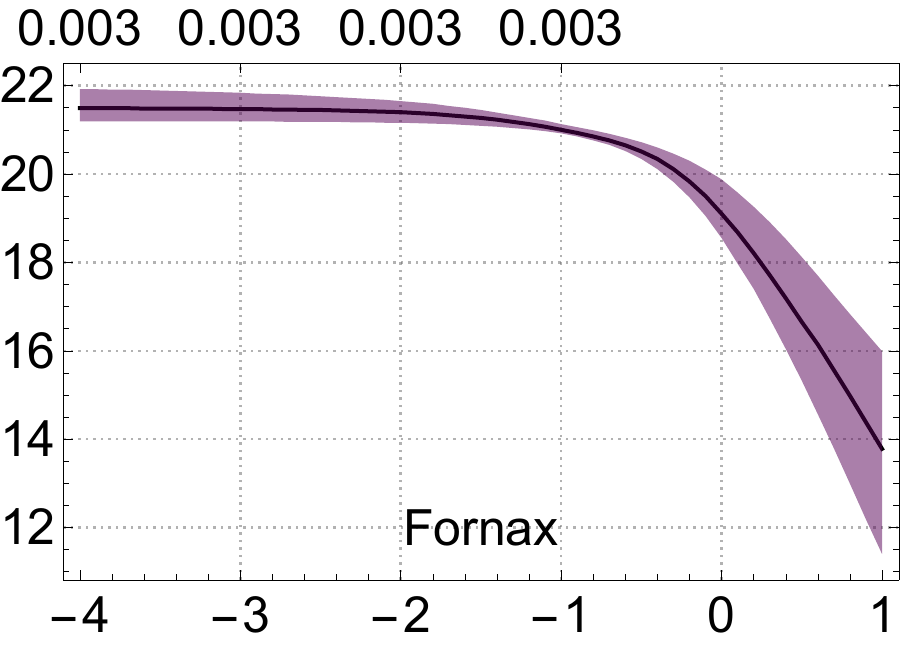}
\includegraphics[width=45mm]{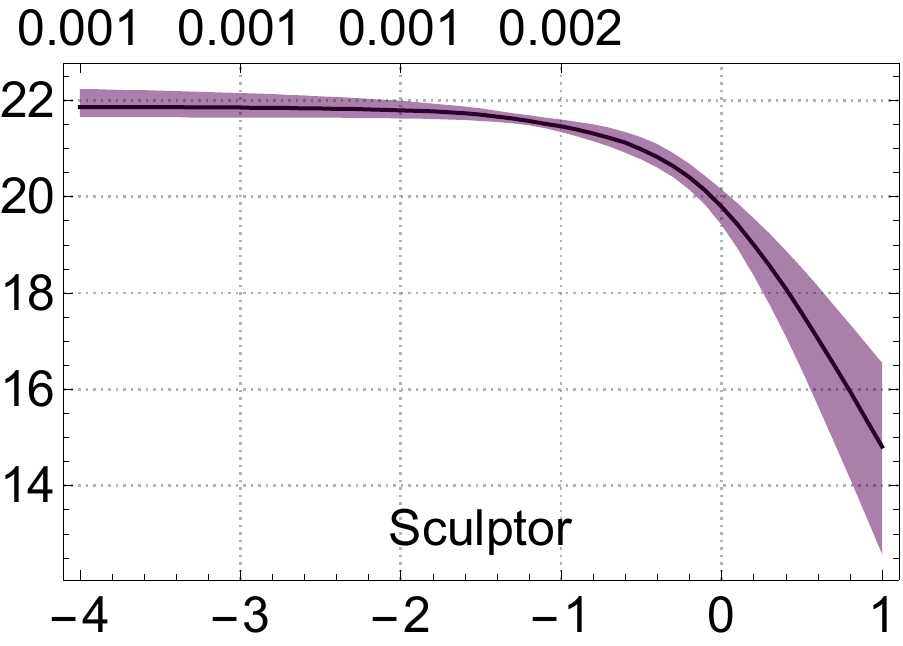}
\includegraphics[width=45mm]{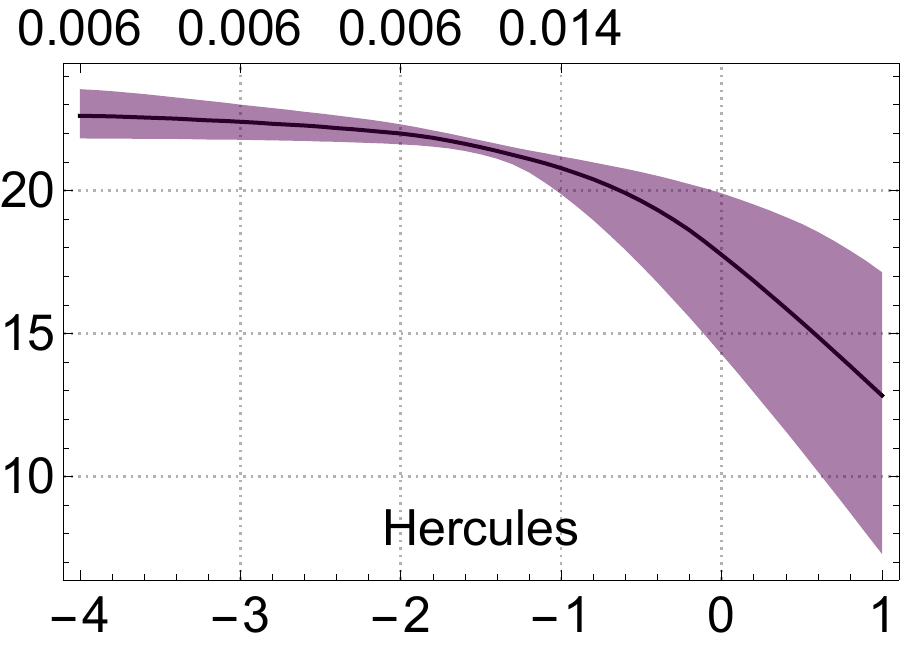}
\includegraphics[width=45mm]{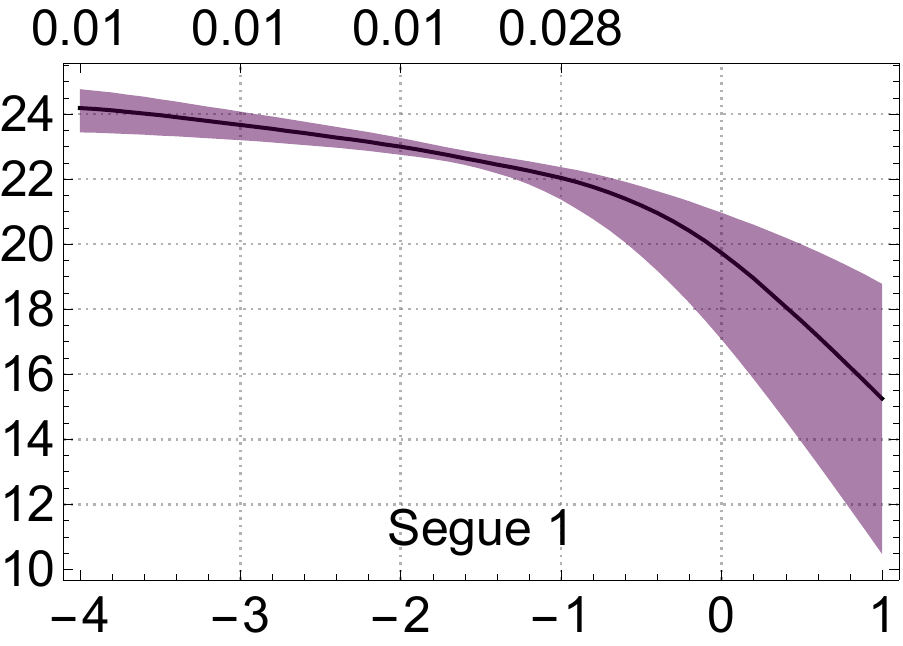}
\end{center}
\caption{Same as Fig.\ref{fig:1} and \ref{fig:2} but for different dSphs. For the labels of the axes see Fig.\ref{fig:1}.} \label{fig:3}
\end{figure}

\begin{figure}[!t]
\begin{center}  
\includegraphics[width=45mm]{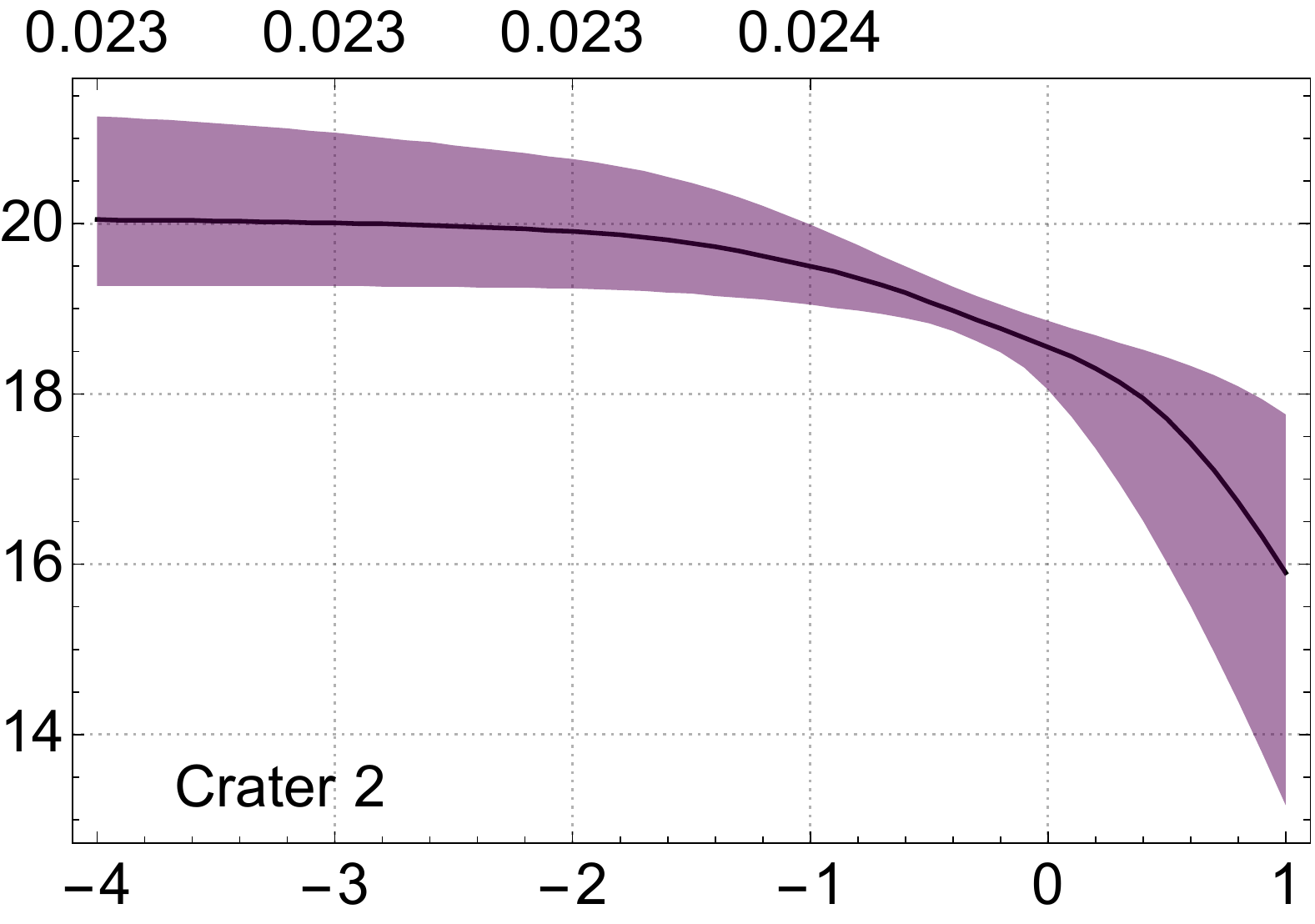}
\includegraphics[width=45mm]{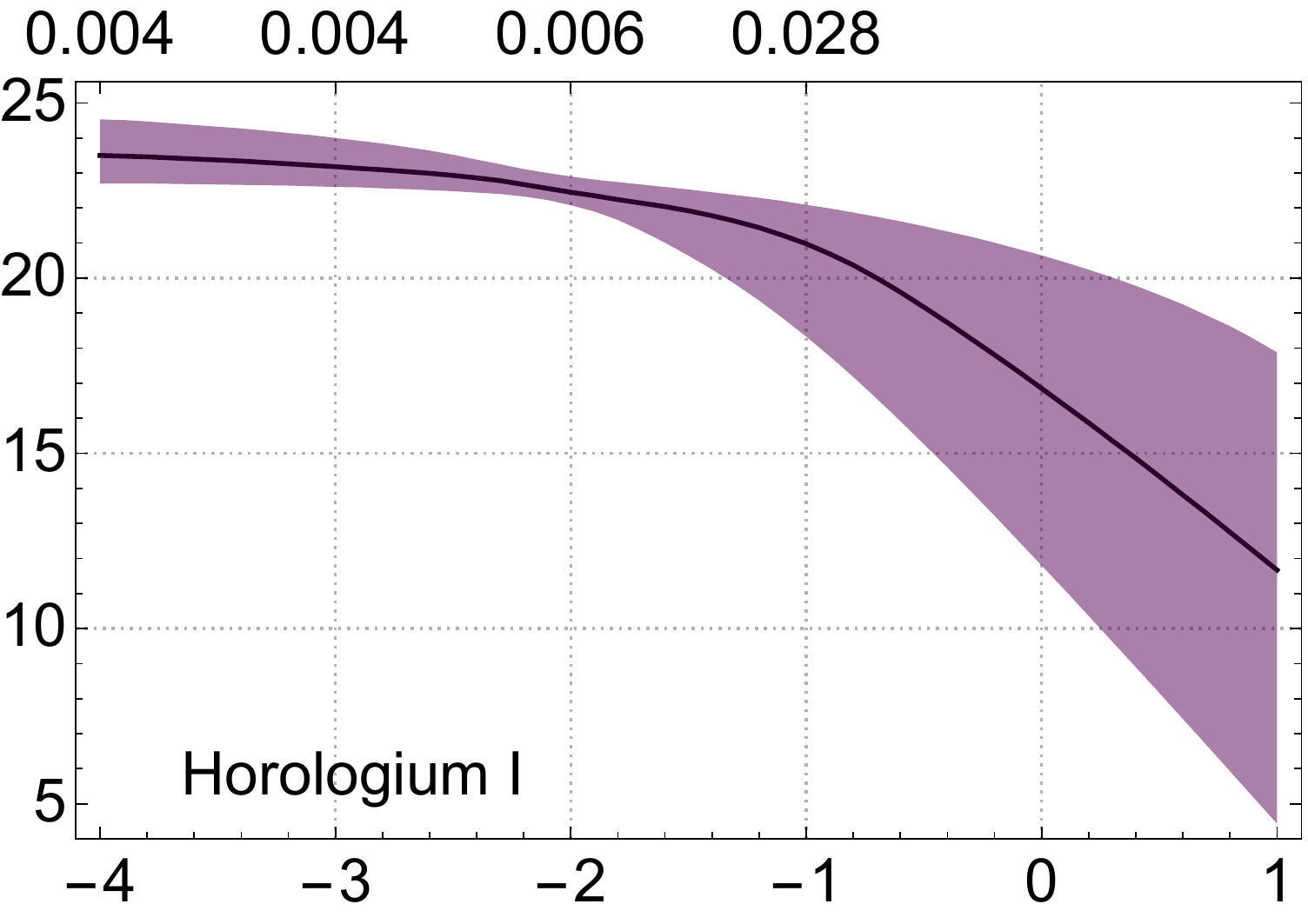}
\includegraphics[width=45mm]{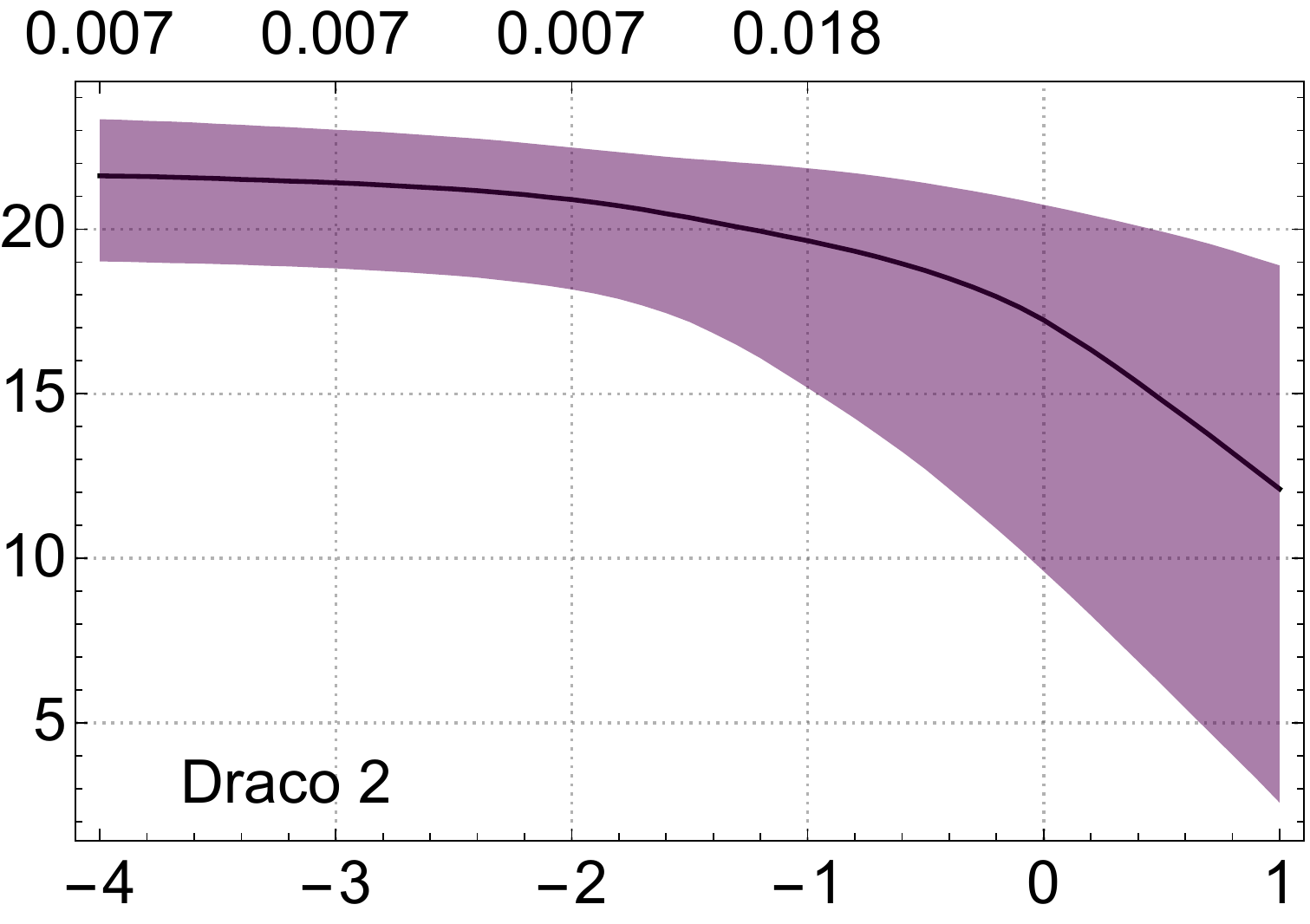}
\includegraphics[width=45mm]{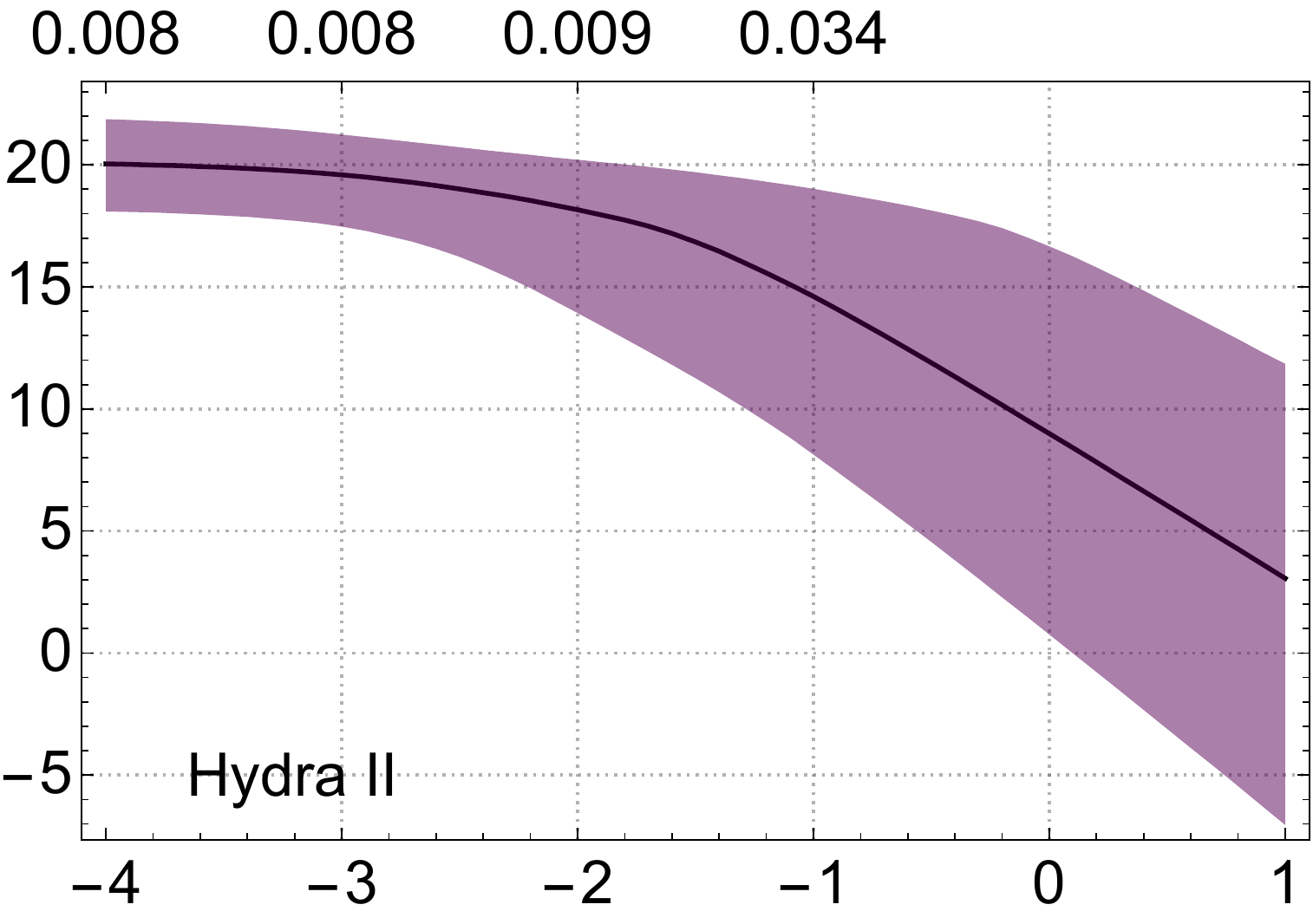}
\includegraphics[width=45mm]{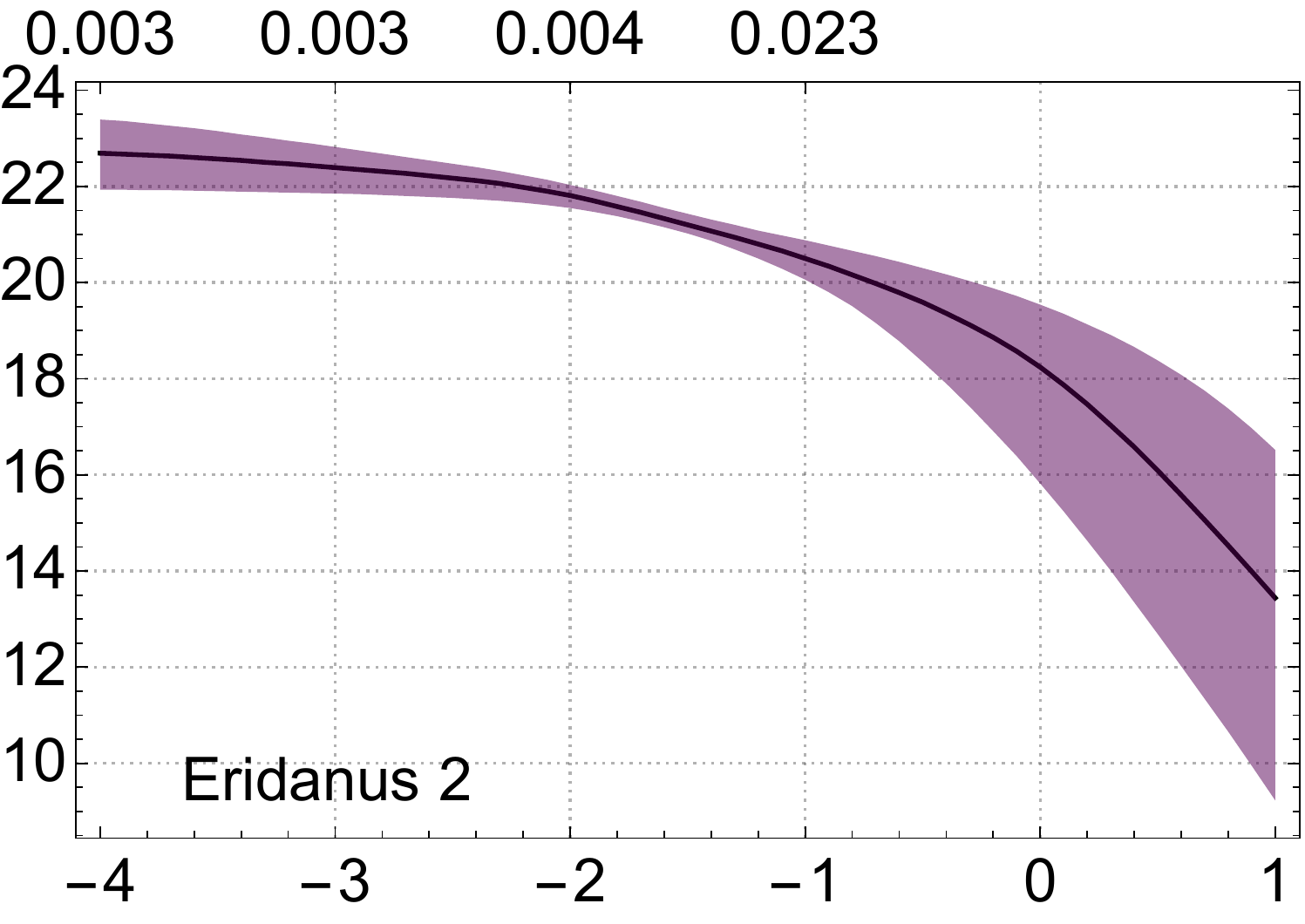}
\includegraphics[width=45mm]{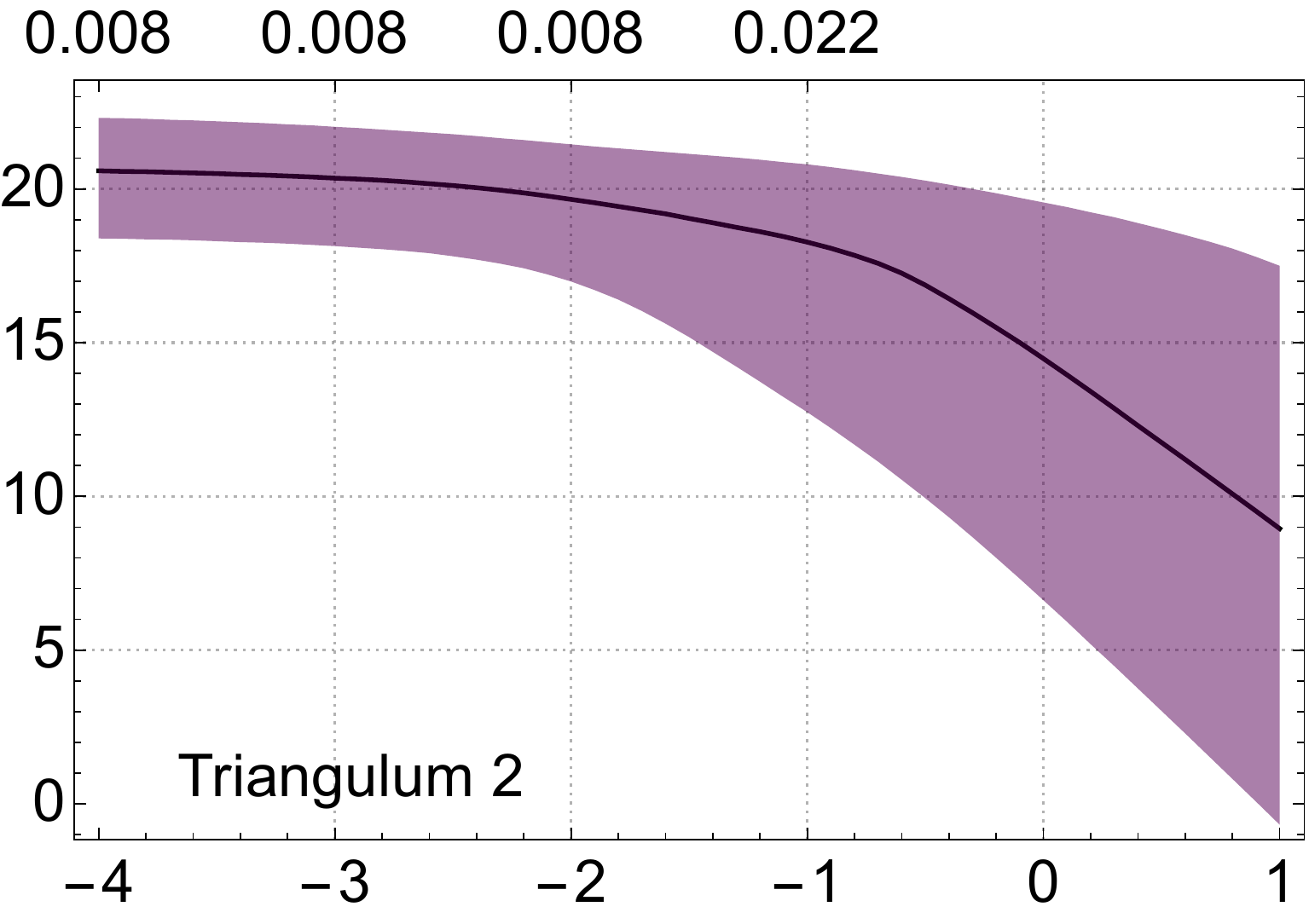}
\includegraphics[width=45mm]{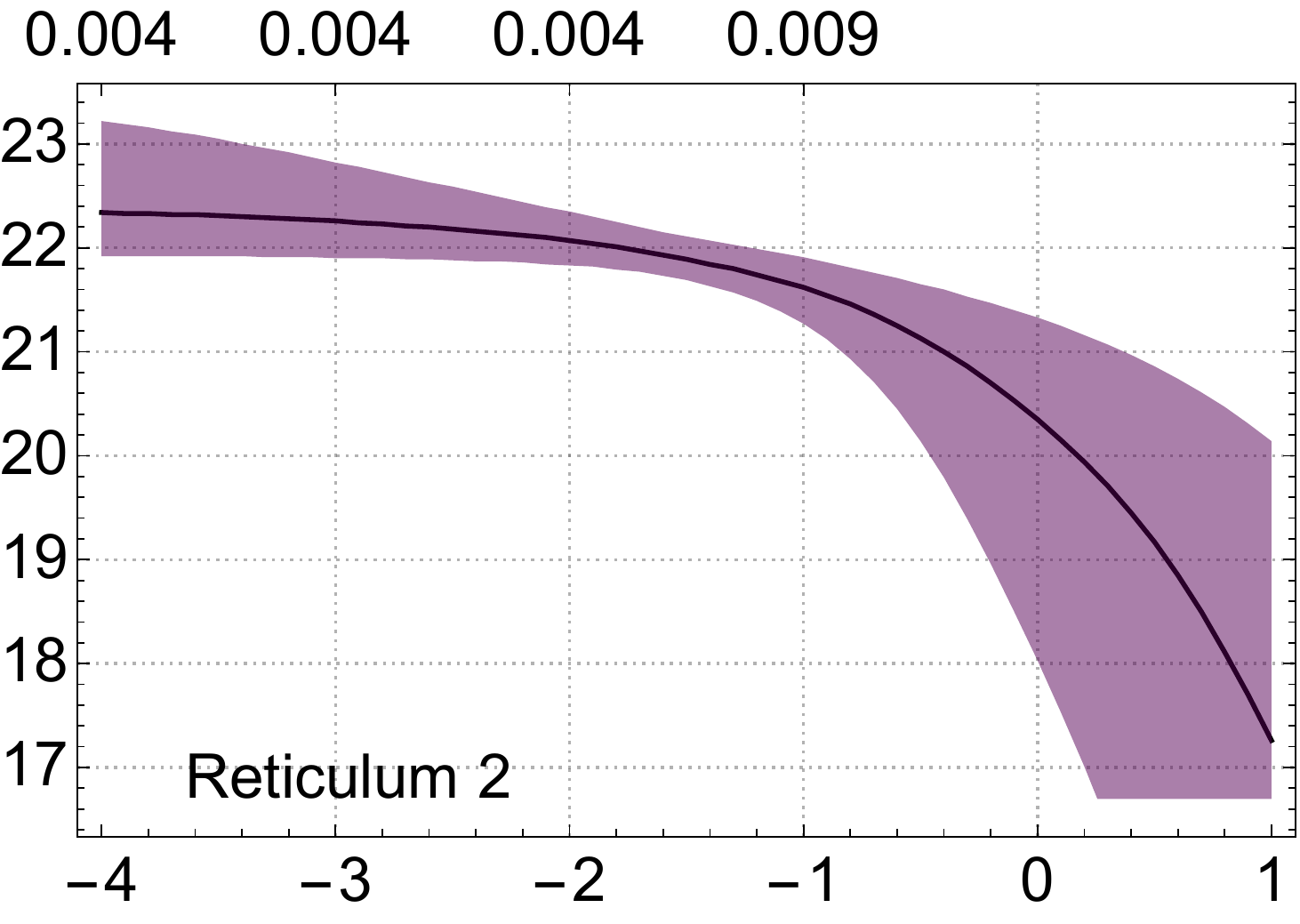}
\includegraphics[width=45mm]{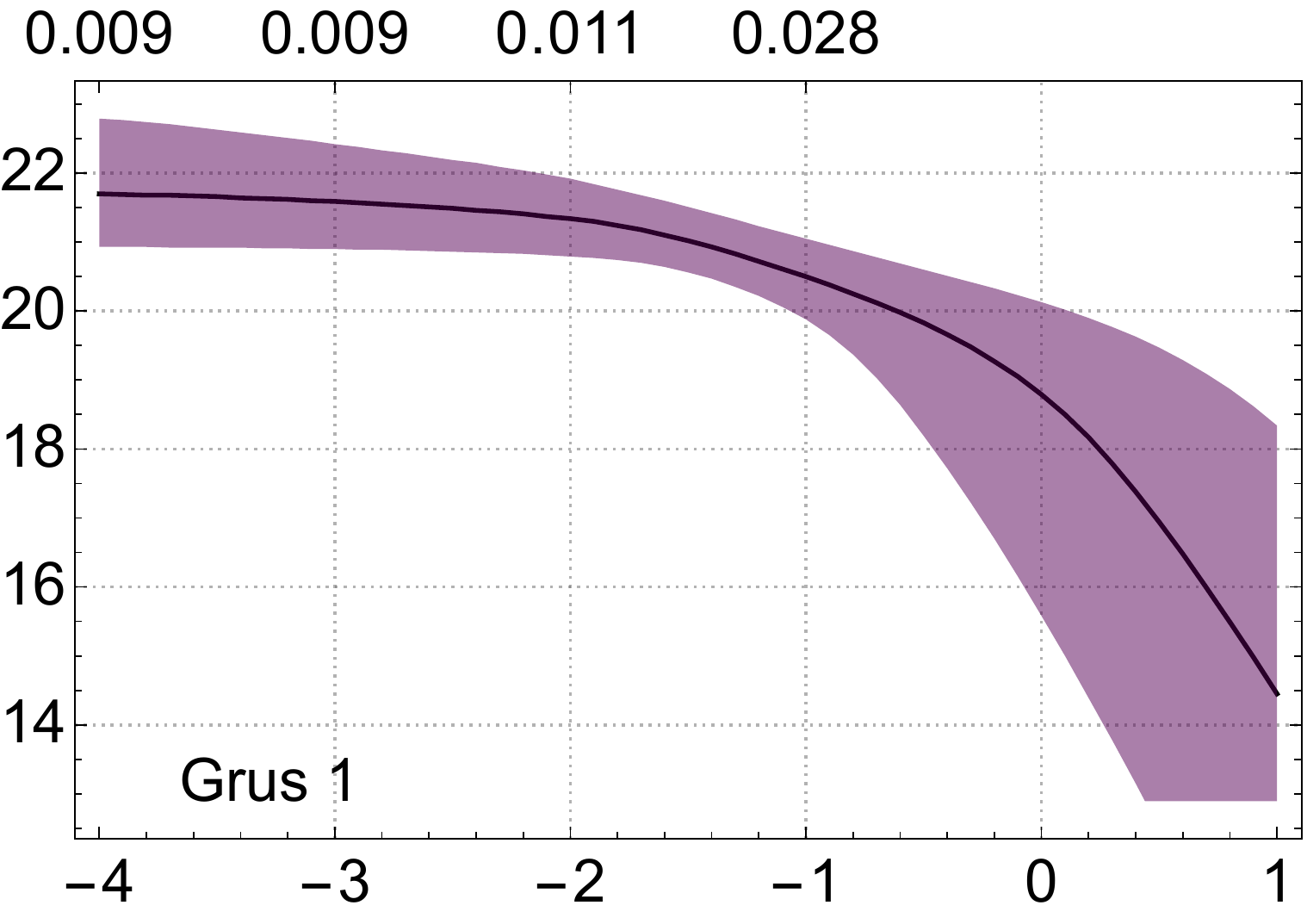}
\includegraphics[width=45mm]{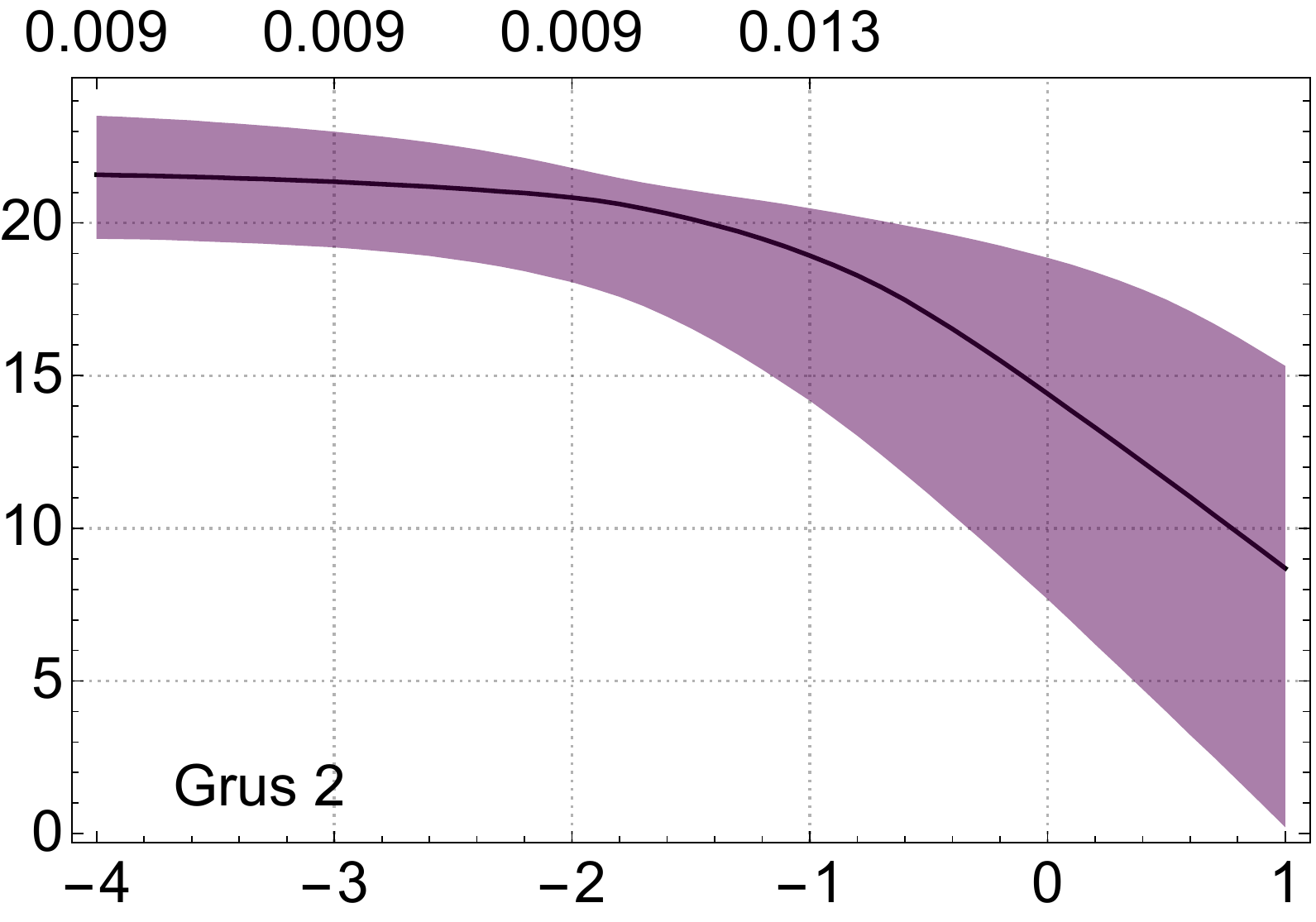}
\includegraphics[width=45mm]{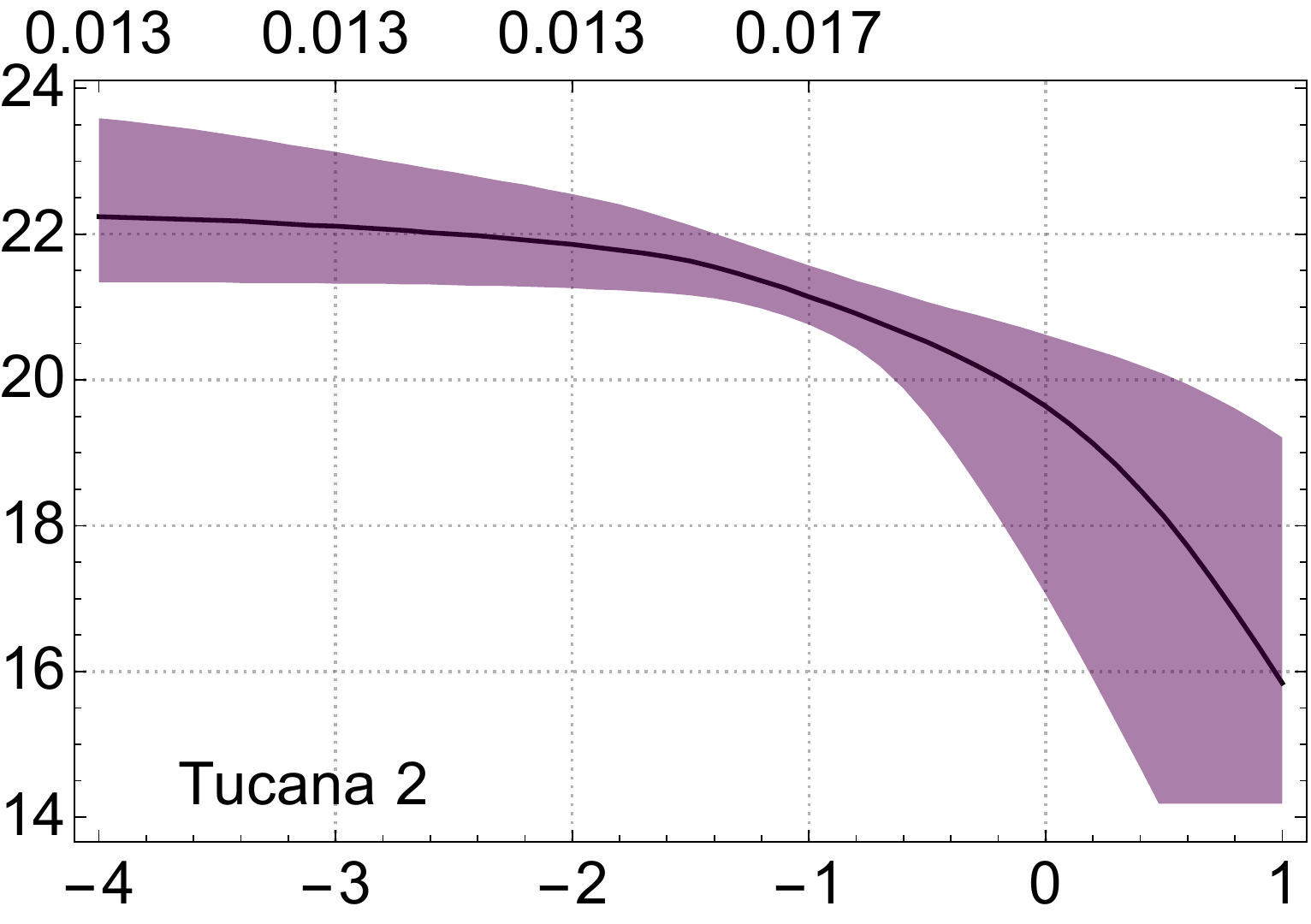}
\includegraphics[width=45mm]{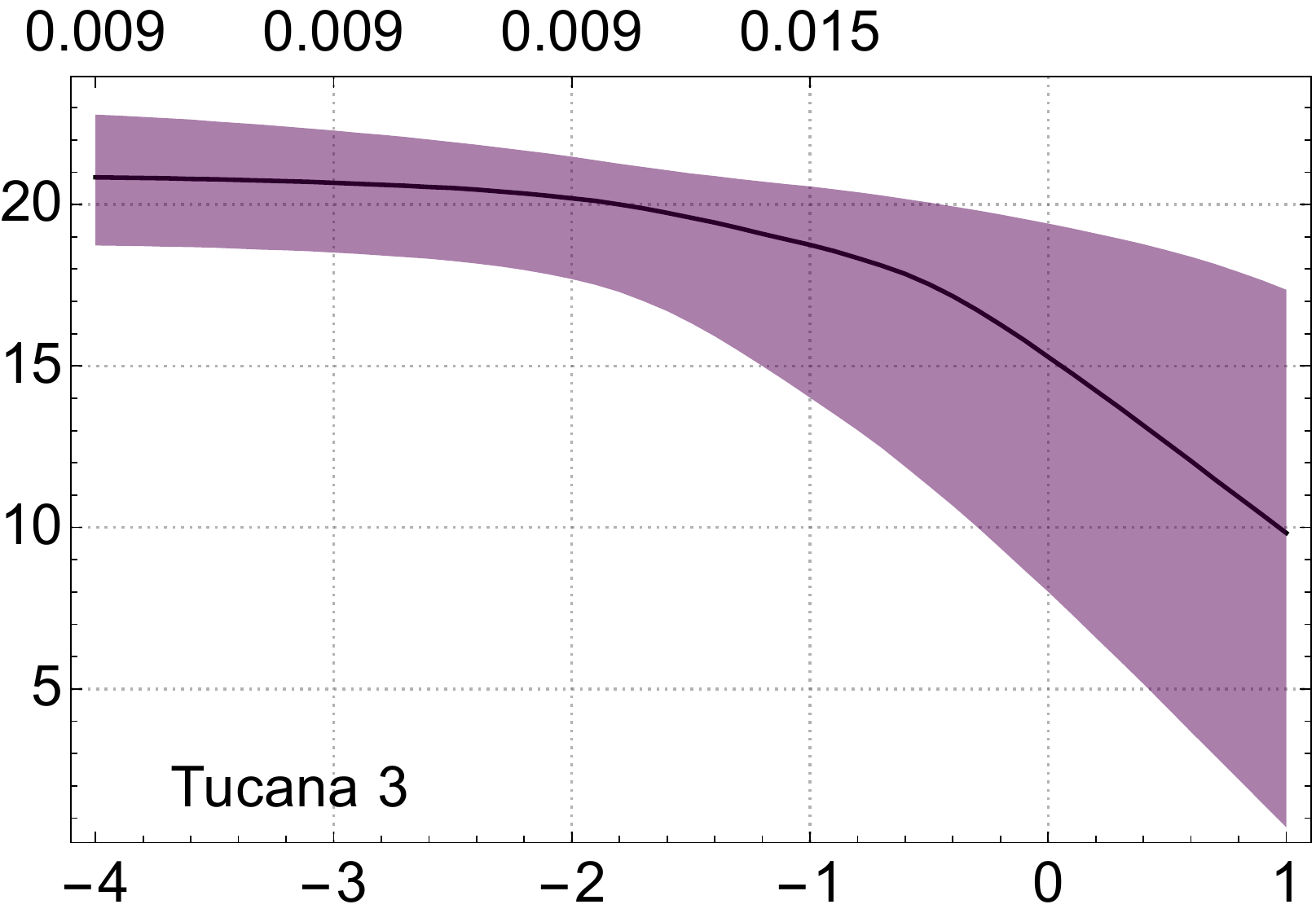}
\includegraphics[width=45mm]{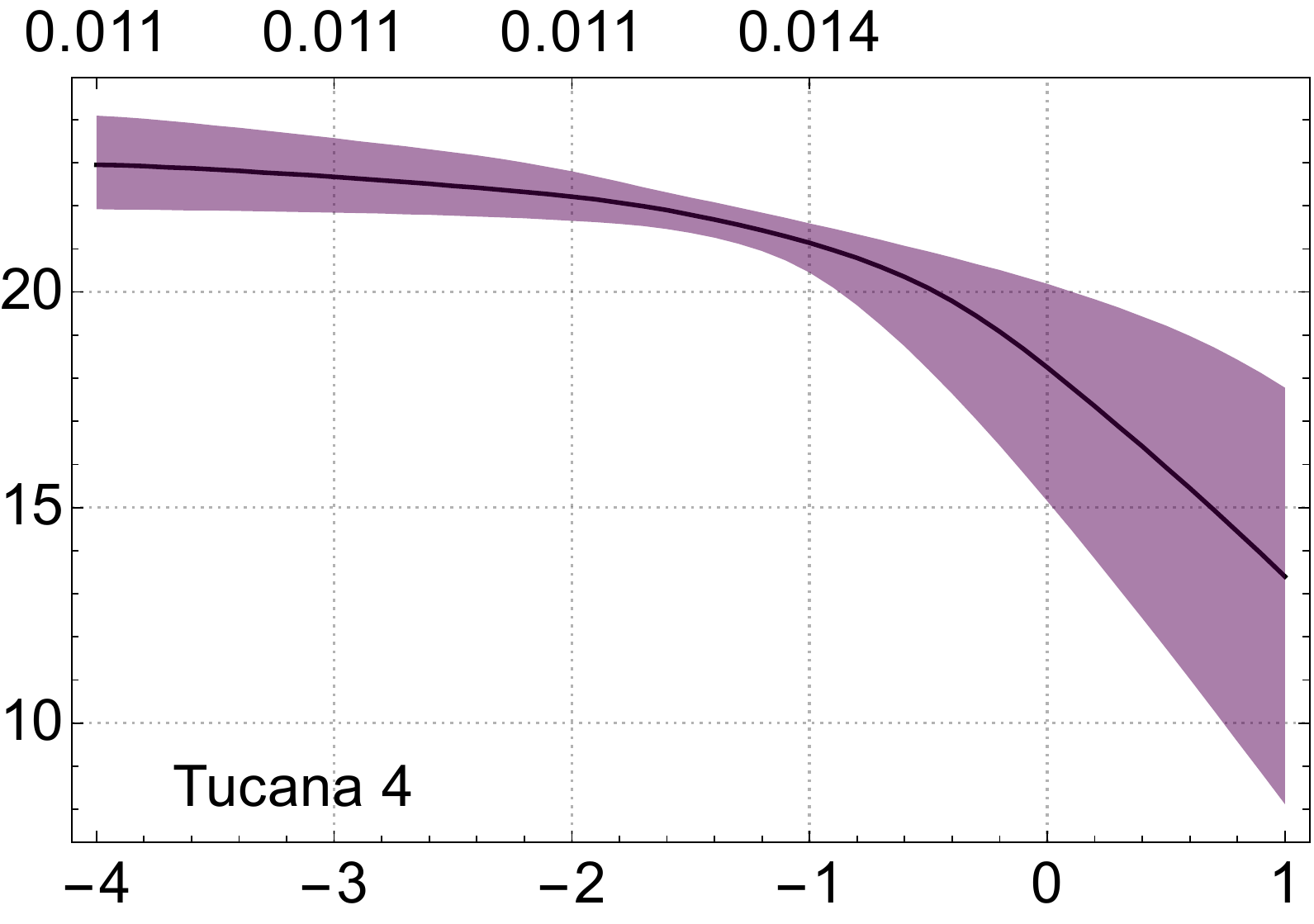}
\includegraphics[width=45mm]{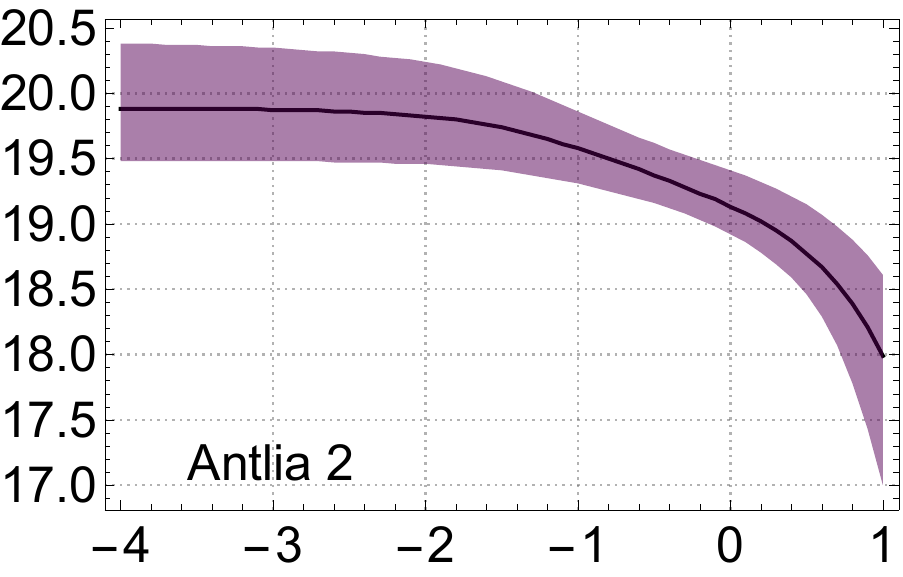}
\end{center}
\caption{Same as Figs.\,\ref{fig:1},\,\ref{fig:2} and \ref{fig:3} but for different dSphs. For the labels of the axes, see Fig.~\ref{fig:1}.
}
 \label{fig:4}
\end{figure}

Then we can factorize the integral to obtain 
\beq
\laq{dsph}
\frac{d^2 \F_{\g,i=\rm dSph}}{d E_\g d \Omega }\simeq \frac{\partial_\Omega D}{4\pi }\frac{\G_{\f}}{m_\f} F_{\rm rest}[\frac{E_\gamma}{1-v_i^{\rm rel}}].
\eeq
Here we defined the differential $D$-factor, 
\beq
\partial_\Omega D\equiv \int d s \rho^i_{\f} (s, \Omega) . 
\eeq
This does not depend on the mass and interaction of the DM as long as it is a single component (while we can easily extend the formula to multi-component cold DM given the fraction of the abundance of each component). 
{Sometimes the averaged $\left[\int d\Omega \partial_\Omega D\right]/(\Delta \Omega)$ around the center of the galaxy was used in literature instead of $\partial_\Omega D$. 
The so-called integrated $D$-factor (the denominator) can be obtained from various references~\citep{Combet:2012tt,Geringer-Sameth:2014yza,Bonnivard:2015xpq,Bonnivard:2015tta,Hayashi:2016kcy,Sanders:2016eie,Evans:2016xwx,Hayashi:2018uop,Petac:2018gue}.
This is a good approximation when the field of view is large enough and the resolution is not very good so that we can average the detail distribution of the DM. 
Instead, here we will focus on $\partial_\Omega D$, because 
\begin{itemize}
\item Avoiding visible stars may not allow us to look at the center of the galaxy. 
\item The signal event rate, as well as the uncertainty, depends on the DM distribution, which must not be uniform in a galaxy. 
\end{itemize}
These effects are important for a detector with a small field of view or/and good angular resolution. 

The mass density profile of a dSph is the key input for estimating its $\partial_\Omega D$.
Different techniques have been developed in order to infer the density profiles from stellar kinematic data, such as distribution function modeling, orbit-based methods, as well as Jeans analysis~\citep{Battaglia:2013wqa,2022NatAs...6..659B}. 
In this work, we utilize the mass density profiles estimated by non-spherical mass models based on axisymmetric Jeans equations~\citep{Hayashi:2020jze,Hayashi:2022wnw} to calculate $\partial_\Omega D$ {(Here the Hernquist profile is adopted \citep{1990ApJ...356..359H,1996MNRAS.278..488Z}. See also Refs.\,~\citep{Navarro:1995iw, 2010MNRAS.402...21N, Burkert:1995yz} for  Navarro–Frenk–White, Einasto, and Burkert profiles.)}. They are shown in Figs.~\ref{fig:1}, \ref{fig:2}, \ref{fig:3} and \ref{fig:4} in the angular distance $\log_{10}[r/\rm deg]$ vs the differential $D$-factor $\log_{10}(\partial_\Omega D/ [\rm GeV/cm^2/sr])$ plane. 
The main reason why we adopt the results from non-spherical mass models is that such models can treat two-dimensional distributions of line-of-sight velocity dispersions and thereby mitigating a strong degeneracy between DM density and stellar velocity anisotropy~(see \citep{Hayashi:2020jze} for the detailed methods). 
{The uncertainty in their estimated $\partial_\Omega D$ mainly stems from a strong degeneracy between the shape of the dark matter halo and the stellar velocity anisotropy. This degeneracy results in significant uncertainty in determining the axial ratio of the dark matter halo. Additionally, especially for ultra-faint dwarf galaxies, the insufficient volume of available data also contributes to the main uncertainties in their $\partial_\Omega D$ values.}

{The differential $D$-factor can be useful in the following senses. 
\begin{itemize}
\item In order to make optimal use of the data, we integrate it over the detector's field of view, taking into account the detector efficiency, $P_{\rm det}$. This efficiency could be a function of both angular position and energy and may also include the effect of masking. As a result, we can obtain an optimized $D$-factor as follows:
\beq
D^{\rm optimized}[E] \equiv \int d \Omega P_{\rm det}(\Omega,E){\partial_\Omega D}.
\eeq
\item One can use the differential $D$-factor to distinguish the DM signal from the noise, especially when the detector's field of view is sufficiently large and the angular resolution is high, e.g., MUSE~\citep{bacon2017muse}. 
We can conduct such an analysis by correlating the angular distribution of the detected photons with the expected distribution of DM. This allows us to ``see" the distribution of DM.
\item We may also use the differential $D$-factor without the angular integral as a good approximation when the field of view is small, which will be discussed soon. 
\end{itemize}
}

In Figs. \ref{fig:1}, \ref{fig:2}, \ref{fig:3} and \ref{fig:4}, we also show the typical {angular distance in between stars} 
defined by 
\beq
\bar d_{\rm stars}\equiv \Sigma^{-1/2}_N[r]
\eeq
with $\Sigma_N$ being the star number density. 
For Figs.~\ref{fig:1}, \ref{fig:2}, \ref{fig:3} we estimated it by using the King Models 
with the data from Ref.\,\citep{Mu_oz_2018} where they estimated
the stellar number density profiles of the galactic dwarf satellites using the photometric data brighter than $g\simeq25.6$~mag from MegaCam wide field imagers on the 3.6~m Canada–France–Hawaii Telescope and the 6.5m Magellan-Clay telescope. 
For Fig.\,\ref{fig:4}, 
the star distance is estimated by using the number density or the fit from Ref.\,\citep{10.1093/mnras/stw733} for Crater 2,  \citep{DES:2015zwj} for Grus 2, Tucana 3, and Tucana 4, \citep{Koposov:2015cua} for Horologium I, Reticulum 2, Eridanus 2, Tucana 2, and Grus 1, \citep{Laevens:2015kla} for Draco 2,  \citep{Martin:2015xla} for Hydra II and \citep{Laevens:2015una} for Triangulum 2. We cannot find the star density for Antlia 2, and we only show $\partial_\Omega D.$ 
We plot the value on the top axis corresponding to $r=10^{-4},10^{-3},$ $10^{-2},10^{-1}$\,deg. We note that the estimation for the star angular distance may be too rough at $r=10^{-1}$\,deg. 
From the typical angular distance between visible stars, we can check how good an angular resolution can benefit.\footnote{We also recommend checking the actual data (e.g., Fig.~\ref{fig:stardistance}) when the resolution is close to the typical star distance.}
}

\section{eV axion search with IRCS in Subaru telescope}

\subsection{Setup and Motivation}
\paragraph{Setup}
Let us consider the DM, $\phi$, decaying into the two particles, including a photon. 
The decay rate is given as $\Gamma_{\phi}.$ 
Although our discussion does not rely on it, for concreteness, let us consider that the DM is an ALP. Then, the Lagrangian is given by 
\beq
{\cal L}=- \frac{g_{\f \g\g} }{4 }\f  F_{\mu\n}\tl{F}^{\m\n}, 
\eeq
with $F_{\m\n}$  ($\tl{F}^{\m\n}$) being the field strength of the photon (it's dual),  and $g_{\f \g\g} $ the photon coupling. 
Namely, the ALP can decay into two photons.  
In the following, we mostly focus on this possibility.
We assume that $\f$ is the cold DM and is, therefore, non-relativistic. 
The resulting photon from the decay, $\f\to \g\g,$ has the energy $E_\g = m_\f/2$ if $\f$ is at rest. Here  $m_\f$ is the mass of the ALP. 
The decay rate of the ALP at rest can be estimated as 
\beq
\G_\f= \frac{g_{\f\g\g}^2}{64\pi} m_\f^3. 
\eeq

\paragraph{Motivation}In particular, in the parameter region $$
m_\phi \sim \,{\rm eV}, ~~~~ g_{\phi \gamma\gamma} \sim 10^{-11}\text{-}10^{-10} {\rm GeV}^{-1}
$$
 there are multiple observational implications: 
 \begin{description}
 \item[1] The tension between theory and observation of cooling of horizontal-branching stars can be alleviated if light enough particle with $g_{\f \g\g }$ in the range is present~\citep{Ayala:2014pea}; 
 \item[2] The gap between the analysis of the measured anisotropy of the cosmic infrared background and the theoretical prediction can be explained by the fact that such DMs decay into photons~\citep{Gong:2015hke,Caputo:2020msf}; 
 \item[3] The bump feature of the cosmic $\gamma$-ray spectrum near TeV can be explained by the annihilation with a photon from the DM decay into the electron and positron~\citep{Korochkin:2019qpe}.
 \end{description}
Although these observational grounds do not rigorously establish the existence of such DM due to astronomical uncertainty, it is notable that the independent physics is relevant around the similar parameter region. 
We also comment that there are constraints from the cooling of the horizontal branch stars relevant to {\bf 1} \citep{Dolan:2022kul} (See also Refs.\,\citep{Raffelt:1985nk,Raffelt:1987yu,Raffelt:1996wa, Ayala:2014pea, Straniero:2015nvc, Giannotti:2015kwo, Carenza:2020zil}) $g_{\f\g\g}\lesssim 4.7\times 10^{-11}\GEV^{-1}.$ 
This is applied if $m_\f\lesssim 10\KEV$. For the eV ALP DM, this is one of the most stringent bounds, and thus it is important if an ALP DM search reaches beyond this bound. Also, there are bounds from the anisotropic cosmic infrared background and the attenuation of the cosmic $\gamma$-ray spectrum relevant to {\bf 2 \AND 3} for ALP mass $m_\f \gtrsim 10\,$eV~\citep{Nakayama:2022jza,Carenza:2023qxh}.
\\

Theoretically, the eV range DM has been the leading candidate of the DM half-century ago, known as hot DM. 
This is because if the DM is produced, such as it is in thermal equilibrium with the standard model plasma, the DM mass range is predicted to be around eV to match the measured DM abundance. 
A similar production of the DM with the mass in the eV range can still have the DM consistent with the $\Lambda$CDM model due to a bose-enhancement effect if the DM is a boson~\citep{Yin:2023jjj}.

 Alternatively, ALPs, with masses near eV, are also predicted theoretically from the hadronic axion window scenario~\citep{Chang:1993gm, Moroi:1998qs} and the ALP miracle scenario~\citep{Daido:2017wwb,Daido:2017tbr}.
They both predict special relations between the photon coupling and ALP mass. 
The first special relationship is 
\beq
\laq{QCDaxion}
g_{\phi \g\g}\simeq \(1.6-400\)\times10^{-11}\GEV^{-1} \(\frac{m_\f}{2\EV}\) {~~[\text{QCD axion}]}.
\eeq 
This is the case that $\f$ is the QCD axion~\citep{Peccei:1977hh,Peccei:1977ur,Weinberg:1977ma,Wilczek:1977pj, GrillidiCortona:2015jxo} that solves the strong CP problem of the standard model. The lower limit of the axion-photon coupling is from the hadronic axion window~\citep{Chang:1993gm, Moroi:1998qs}, in which the photon coupling is accidentally canceled with the model-dependent parameter $E/N=2$~\citep{Moroi:1998qs}.  
More precisely, we take the lowest value to be the hadronic uncertainty~\citep{GrillidiCortona:2015jxo} when this accidental cancellation happens. 
We take $(E/N)_{\rm max}=12$. 
In this scenario, the axion mass points to the eV range, bounded by the constraints relevant to the SN1987A neutrino burst (See also \citep{Chang:2018rso,Bar:2019ifz}). The hot DM bound can be alleviated if we consider some more specific production~\citep{Moroi:2020has, Moroi:2020bkq} with low reheating temperature~\citep{Grin:2007yg,Carenza:2021ebx}. Alternatively, we can build a model to have a cooling of the thermal plasma together with the aforementioned ``thermal" production with the Bose enhancement to explain the cold DM~\citep{Yin:2023jjj}.

The second relation is from the hypothesis that the ALP serves as both DM and inflaton~\citep{Daido:2017wwb, Daido:2017tbr} (see also \citep{IAXO:2019mpb}). For the primordial density perturbation to be explained by the ALP inflation, we get
\beq \laq{ALP}
g_{\phi \g\g}\simeq c_\g 10^{-11}\GEV^{-1}\(\frac{2\EV }{m_\f } \)^{1/2} [{\rm ALP=DM=inflaton}]
\eeq
Here, $c_\g$ is a model-dependent parameter, and we consider the range $c_\g=0.1-10$ as the natural region. Additionally, successful cosmology without further assumptions predicts the ALP mass to be in the range $m_\f=\O(0.01-1)\EV$, based on constraints from small-scale structures, successful reheating, and the production of the correct DM abundance. 
The existence of a viable parameter region that satisfies various constraints is non-trivial, and the minimal scenario is known as the ALP miracle (See also cf. Refs.\,\citep{Takahashi:2021tff,Takahashi:2020uio} and Ref.\,\citep{Takahashi:2023vhv} where the strong CP problem of the standard model can also be explained).

The ALP cold DM in our Universe must be non-relativistic and naturally decays into two photons. 
In particular, the dSphs, our focus, have large central mass-to-light ratios of $\sim \O(10-100)$, and small velocity dispersion, $\lesssim 10$ km/s, as we have mentioned. 
The resulting decay photon has a line-like spectrum for the detector energy resolution of interest.
Therefore searching for the line-like photon from the dSphs in the eV range (infrared and optical photons) should be well-motivated.

\subsection{ Searching for ALP DM in Draco and Ursa Major II by Subaru-IRCS}

Let us discuss one example for using our derived $\partial_\Omega D$ for DM detection.  We propose to search for eV range DM by using an infrared spectrograph, IRCS, in the Subaru Telescope. Here our main targets are Draco~{(central sky coordinates $\alpha_{2000}=$17:20:12, $\delta_{2000}=+$57:54:55, distance~$D_\odot=76\pm6$~kpc)} and Ursa Major II~{($\alpha_{2000}=$08:51:30, $\delta_{2000}=+$63:07:48, $D_\odot=32\pm4$~kpc)}.
As discussed in Ref.\,\citep{Bessho:2022yyu}, the sky noise can be well suppressed when the detector resolution is high enough in the case the target has small velocity dispersion for the DM, and in the case, it decays into two body. 
There are two kinds of sky noise. The first one can be estimated following the IPAC IRSA model (see, e.g., Refs.,\citep{Arendt:1998aj, Schlegel:1997yv, Zubko:2003eg, 1997IPAC, 1998IPAC, Brandt:2011ka}).\footnote{\url{https://irsa.ipac.caltech.edu/applications/BackgroundModel/}} 
In particular, one of the most severe contributions is the zodiacal light which depends on the relative positions of the Earth, the Sun, and the target. Thus it depends on the schedule of the observation.
We have checked that our targets can be observed on dates consistent with the observational schedule of the Subaru telescope, and on some dates, the sky noise (suppressed by the high spectral resolution $R=\O(10^4)$) is much less significant compared to the line-like signal. Since we only plan to observe the targets on those dates, we neglect the contribution from the sky noise estimated by the background model in the following.

\begin{figure}[!t]
\begin{center}  
\includegraphics[scale=0.43]{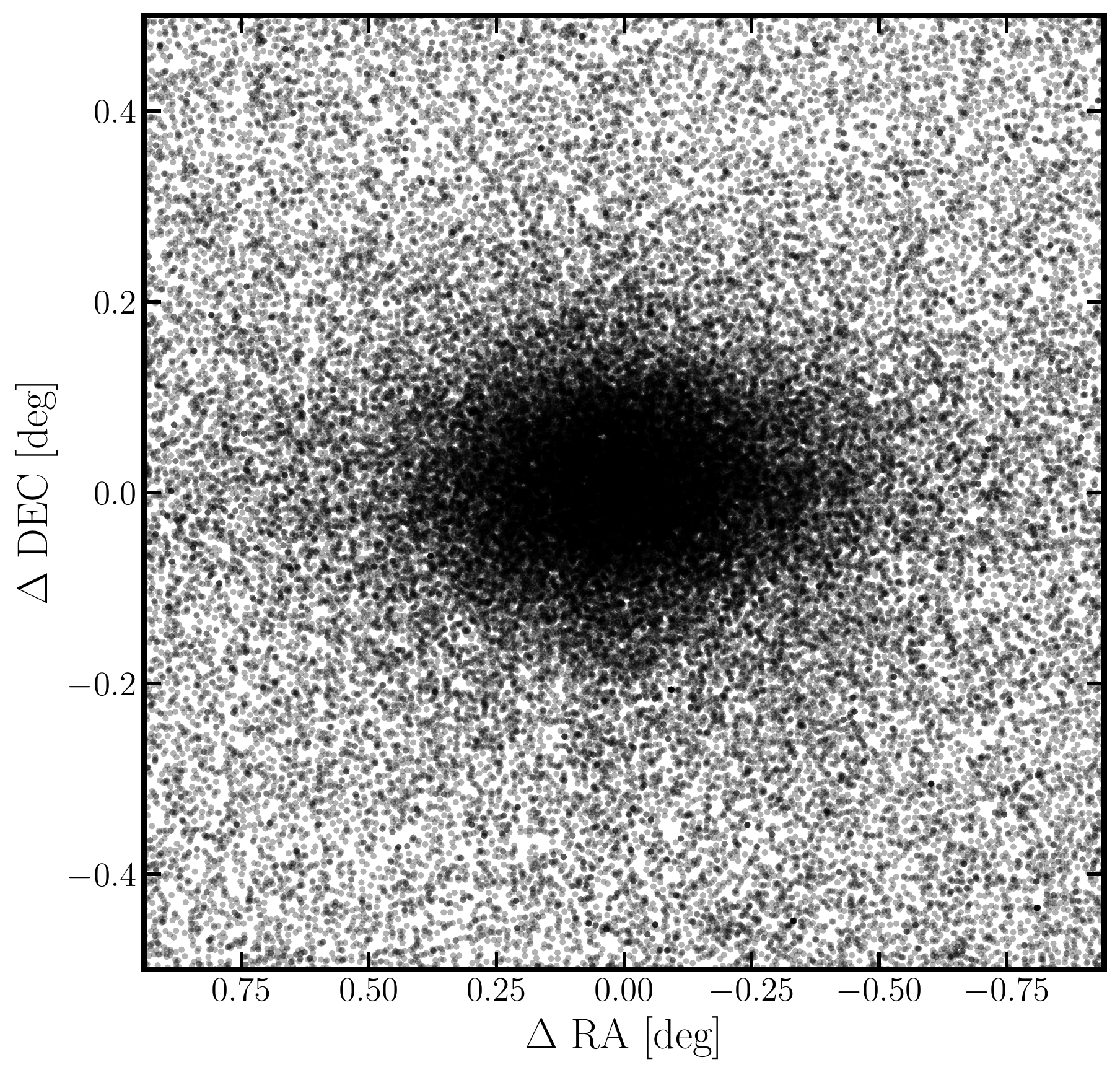}
\includegraphics[scale=0.43]{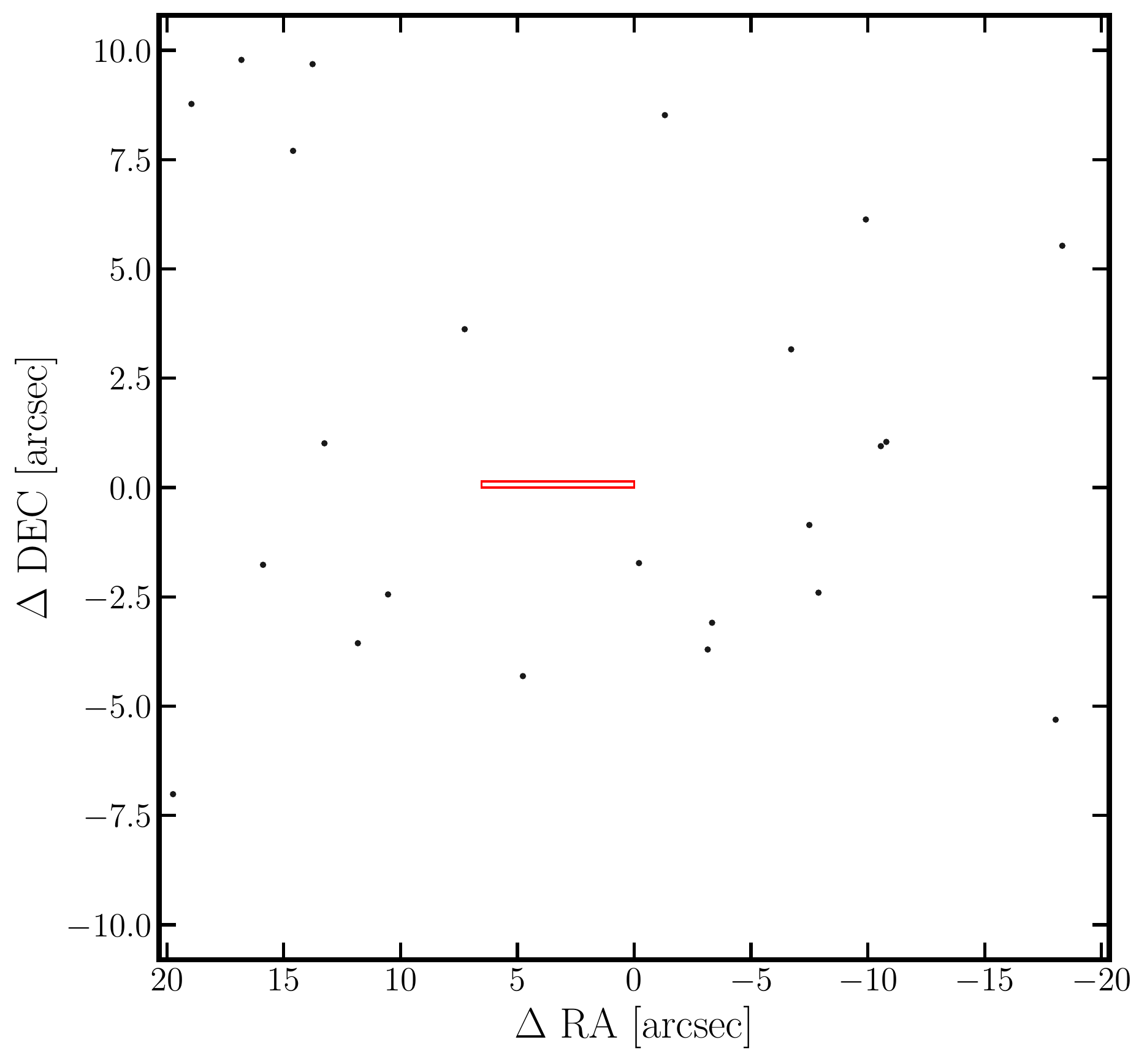}
\end{center}
\caption{Star distribution of Draco observed by {MegaCam} installed on Canada-France-Hawaii telescope (the photometric data taken from \citep{Mu_oz_2018a}). The zero values of each axis indicate the central sky coordinate of Draco~($\alpha_{2000}=$17:20:12, $\delta_{2000}=+$57:54:55). The left and right panels show the spatial distribution of the stars within 1~deg$^2$ and 40~arcsec$^2$, respectively.
The red rectangle in the right panel denotes the slit size of the IRCS spectroscopy for H-band.} \label{fig:stardistance}
\end{figure}

The second is the light from visible stars in the field of view of the target dSphs. 
Indeed, the light can significantly hamper the detection of our target DM particle. Here, we discuss the extent to which stars are crowded in the centers of dSphs for DM detection by comparing them with the field of view of IRCS.
In Fig. \ref{fig:1} (see also Ref.~\citep{Mu_oz_2018}) we see the typical distance is $\O(10), \O(100)$arcsec (the star number densities are $\sim0.01$~stars/arcsec$^{2}$ and $\sim0.001$~stars/arcsec$^{2}$) at the center of Draco and Ursa~Major~II, respectively. 
Furthermore, it is found from Figure~\ref{fig:stardistance} that the stellar densities are very sparse even in the central region of the dSphs.
Compared to the slit size of IRCS ($\O$(0.1)''$\times\O$(1)'') one can easily avoid seeing the visible star.

\begin{figure}[!t]
\begin{center}  
\includegraphics[width=135mm]{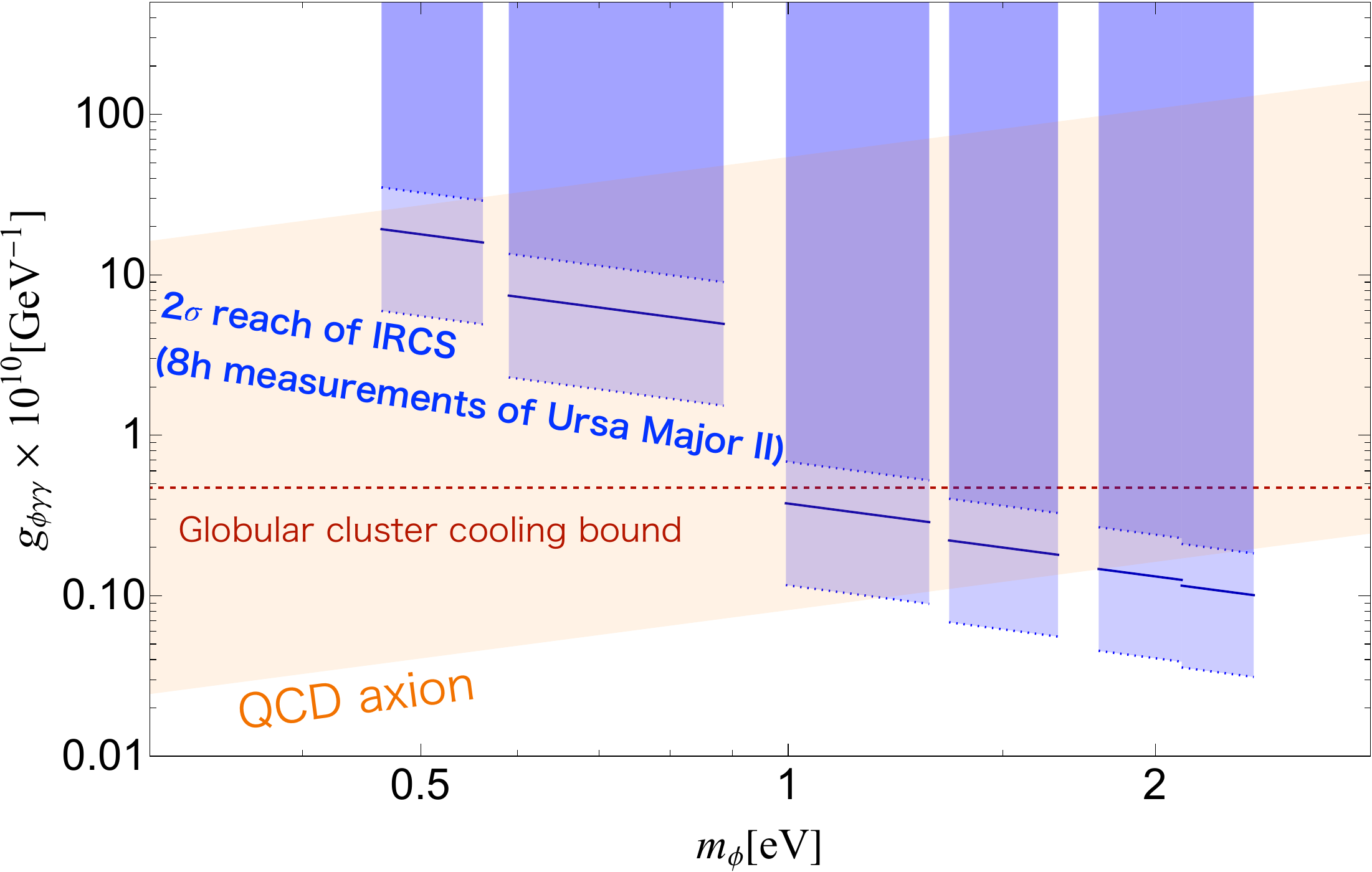}
\includegraphics[width=135mm]{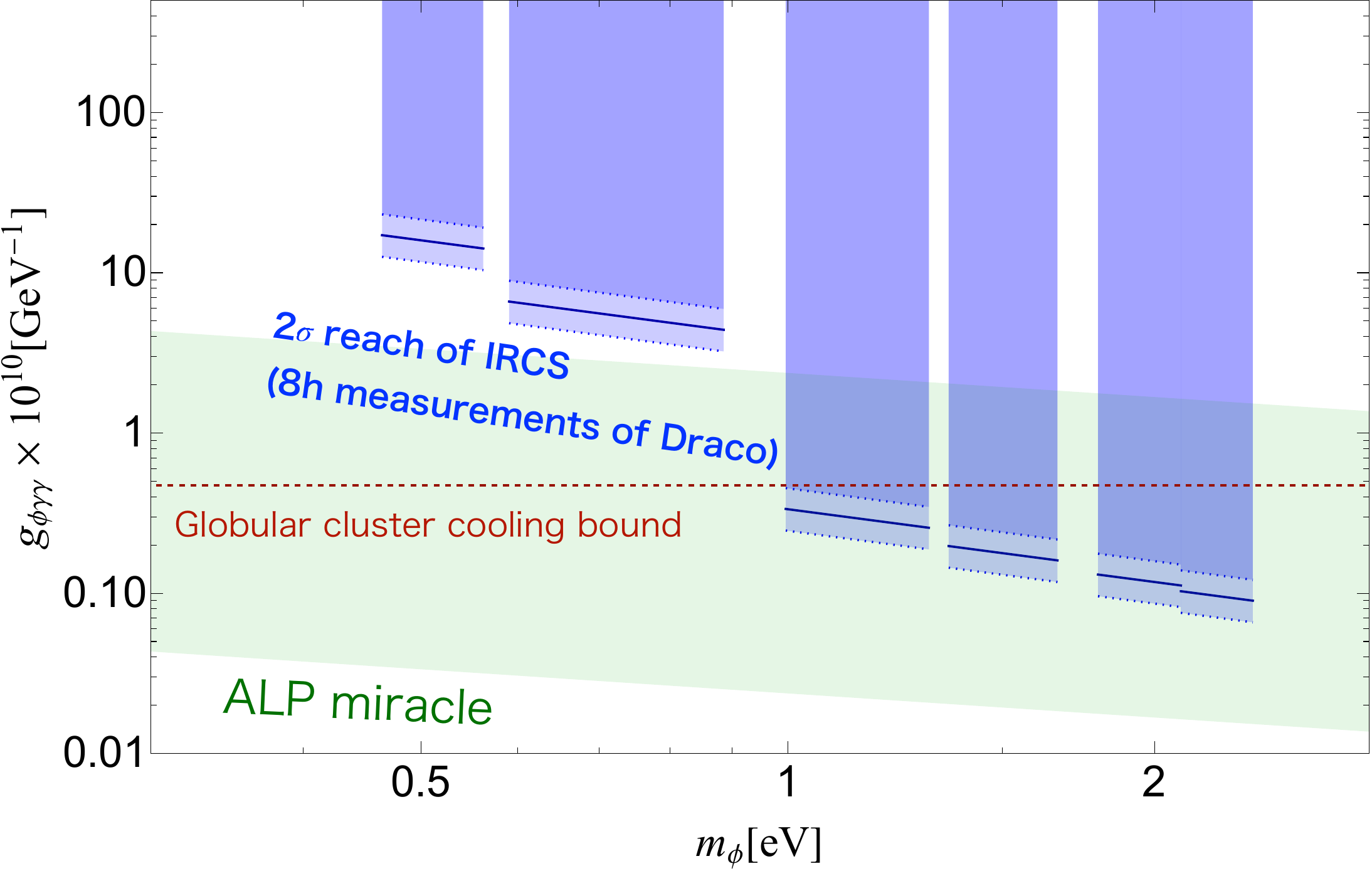}
\end{center}
\caption{The sensitivity reaches of the IRCS for 8 hours observations of Ursa Major II and Draco are shown in the top panel and bottom panel, respectively. From blue bands from  left to right in each panel correspond to M, L, K, H, J, zJ bands. We also denote the 1~$\sigma$ uncertainty from $\partial D/\partial \Omega$.
The stellar cooling bound is shown in the horizontal red dashed line. The orange and green shaded regions are for theoretically well-motivated model predictions.} \label{fig:sensitivity}
\end{figure}

{In the analysis of the differential $D$-factor in this paper following \citep{Hayashi:2020jze,Hayashi:2022wnw}}, both Draco and Ursa~Major~II favor a cusped inner slope of their DM density profiles.
Using their resultant DM density profiles, we calculate $\partial_\Omega D$ distributions for Draco and Ursa~Major~II.
Focusing on the center of the dSphs, the values of $\log_{10}(\partial_\Omega D/(\rm GeV cm^{-2}sr^{-1})= 22.60^{+0.27}_{-0.26}$ for Draco and $= 22.50^{+1.02}_{-0.52}$  for Ursa~Major~II, respectively, at $r\sim \rm arcsec$.
For comparison, the naive average gives $\log_{10}\{(\int d\Omega \partial_\Omega D)/\Delta \Omega /({\rm GeV cm^{-2}sr^{-1})}\}= \log_{10}\{ D[\Delta \Omega]/(\Delta \Omega) /(\rm GeV cm^{-2}sr^{-1})\} \sim 21.07_{-0.16}^{+0.20} \AND 21.45^{+0.38}_{-0.48}$ for Draco and Ursa-Major II, respectively with $\D \Omega= (0.5^\circ)^2 \pi$ as an angular area of the disk around the center.
Therefore we get a factor more than $\O$(10) enhancement than the analysis with the conventional average.
The uncertainty is also very different. They are because by taking the average, we would have the dominant contribution as well as uncertainty to the $D$-factor coming from the DM living at a position far away from the one we look at. 
These high values indicate that the dSphs are the promising targets for detecting eV-scale DM by Subaru-IRCS.

The sensitivities for zJ, J, H, K, L, and M bands are obtained by using the exposure time calculator for IRCS,\footnote{https://www.naoj.org/Observing/Instruments/IRCS/index.html} and are summarized in Table~\ref{tab:data}. 
As a result, by using the differential $D$-factors studied in the previous section, we derive the sensitivity reaches in Fig.~\ref{fig:sensitivity}. 
One finds that $\O$(1) hours' observation can explore the ALP DM in the $1-2\EV$.
We note that it is possible to have the H, K band reaching beyond the star cooling bound within $\O$(1) hours observations 
thanks to the $\O(10)$ enhancement of the flux from our estimation of the differential $D$-factor.

\begin{table}[h]
\caption{Sensitivities in $\rm mag/arcsec^2$ for 8 hours observation by IRCS and parameters used for the estimation. 
The sensitivities are obtained by using the exposure time calculator for IRCS.
}
\label{tab:data} 
\begin{center}
\begin{tabular}{|c|c|c|c|c||c|}
\hline
band & $\lambda_{\rm min}(\mu \rm m)$ & $\lambda_{\rm max} (\mu \rm m)$ & Pixels & R & ${\rm mag}/{\rm arcsec}^2$ \\ \hline
zj & 1.03 & 1.18 & 131 & 20000 & 21.8 \\ \hline
J & 1.18 & 1.38 & 131 & 20000 & 21.6 \\ \hline
H & 1.49 & 1.83 & 195 & 20000 & 21.3 \\ \hline
K & 1.9 & 2.49 & 195 & 20000 & 20.8 \\ \hline
L & 2.8 & 4.2 &  684 & 10000 & 16.3 \\ \hline
M & 4.41 & 5.34 & 684 & 10000 & 14.7 \\ \hline
\end{tabular}
\end{center}
% \begin{tabular}{|c|c|c|c|c|c|}
% \hline
%  band & $\lambda_{\rm min}(\mu \rm m)$ & $\lambda_{\rm max} (\mu \rm m)$ &
%  ${\rm mag}/{\rm sec}^2$ & Pixels & R \\ \hline
%  zj & 1.03 & 1.18 & 21.8
%  % 21.79 
%  & 131 & 20000 \\ \hline
%  J & 1.18 & 1.38 & 
%  21.6
%  %21.61 
%  & 131 & 20000 \\ \hline
%  H & 1.49 & 1.83 & 
%  21.3
% % 21.334 
% & 195 & 20000 \\ \hline
%  K & 1.9 & 2.49 & 20.8
%  % 20.844 
%  & 195 & 20000 \\ \hline
%  L & 2.8 & 4.2 &  16.3
%  %16.2616
%  & 684 & 10000 \\ \hline
%  M & 4.41 & 5.34 & 14.7
% % 14.7117 
%  & 684 & 10000 \\ \hline
% \end{tabular}
\end{table}

\section{Conclusions}
In this study, we conducted a meticulous analysis aimed at indirectly detecting dark matter with a detector that has a high angular resolution or/and a small field of view, which is the case for various spectrograph detectors which can search for light dark matter. 
We proposed to use the differential $D$-factor for the estimation of the signal rate, and we provided it for various dSphs. 
Compared to the conventionally used averaged $D$-factor over a larger area, we found that the resulting dark matter signal can be very different. 
We noticed an enhancement of the dark matter signal rate and a significantly different theoretical uncertainty. They are crucial when we derive a bound or sensitivity reach to a dark matter model. We also provided a new strategy to distinguish the DM signal from the noise, i.e., checking the distribution of the differential $D$-factor from the decay product flux, given that the noise should not have a similar angular distribution of DM. 
This analysis should be viable with detectors having a large enough field of view and good angular resolution.

By concentrating on the decay of dark matter into two particles, one of which is a photon, in the Draco and Ursa Major II dwarf spheroidal galaxies, we demonstrated that an $\O(1)$ nights observation by IRCS in Subaru Telescope yields sensitivity for the decay rate at a world-record level.
In terms of axion-like particles, this surpasses the star cooling bounds in the mass range of 1 eV to 2 eV, aligning with various hints related to the cosmic-infrared background and theoretical predictions of some well-motivated models.

Our study strongly advocates for the actual observations of the corresponding dSphs and offers an applicable approach for the indirect detection of dark matter using high-energy and high-angular-resolution detectors.

\section*{Acknowledgement}
This work was supported by JSPS KAKENHI Grant Nos.  20H05851, 21K20364, 22K14029, and 22H01215 for W.Y., and No. 20H01895, 21K13909, 21H05447, and 23H04009 for K.H.

\bibliographystyle{aasjournal.bst}
\bibliography{ref}

\begin{thebibliography}{}
\expandafter\ifx\csname natexlab\endcsname\relax\def\natexlab#1{#1}\fi
\providecommand{\url}[1]{\href{#1}{#1}}
\providecommand{\dodoi}[1]{doi:~\href{http://doi.org/#1}{\nolinkurl{#1}}}
\providecommand{\doeprint}[1]{\href{http://ascl.net/#1}{\nolinkurl{http://ascl.net/#1}}}
\providecommand{\doarXiv}[1]{\href{https://arxiv.org/abs/#1}{\nolinkurl{https://arxiv.org/abs/#1}}}

\bibitem[{Aalbers {et~al.}(2022)}]{LZ:2022ufs}
Aalbers, J., {et~al.} 2022.
\newblock \doarXiv{2207.03764}

\bibitem[{Abdollahi {et~al.}(2018)}]{Fermi-LAT:2018lqt}
Abdollahi, S., {et~al.} 2018, Science, 362, 1031,
  \dodoi{10.1126/science.aat8123}

\bibitem[{Acharya {et~al.}(2018)}]{CTAConsortium:2017dvg}
Acharya, B.~S., {et~al.} 2018, {Science with the Cherenkov Telescope Array}
  (WSP), \dodoi{10.1142/10986}

\bibitem[{Ackermann {et~al.}(2012)}]{Fermi-LAT:2012edv}
Ackermann, M., {et~al.} 2012, Astrophys. J., 750, 3,
  \dodoi{10.1088/0004-637X/750/1/3}

\bibitem[{Aprile {et~al.}(2020)}]{XENON:2020kmp}
Aprile, E., {et~al.} 2020, JCAP, 11, 031, \dodoi{10.1088/1475-7516/2020/11/031}

\bibitem[{Arendt {et~al.}(1998)}]{Arendt:1998aj}
Arendt, R.~G., {et~al.} 1998, Astrophys. J., 508, 74, \dodoi{10.1086/306381}

\bibitem[{Armengaud {et~al.}(2019)}]{IAXO:2019mpb}
Armengaud, E., {et~al.} 2019, JCAP, 06, 047,
  \dodoi{10.1088/1475-7516/2019/06/047}

\bibitem[{Ayala {et~al.}(2014)Ayala, Dom\'\i{}nguez, Giannotti, Mirizzi, \&
  Straniero}]{Ayala:2014pea}
Ayala, A., Dom\'\i{}nguez, I., Giannotti, M., Mirizzi, A., \& Straniero, O.
  2014, Phys. Rev. Lett., 113, 191302, \dodoi{10.1103/PhysRevLett.113.191302}

\bibitem[{Bacon {et~al.}(2017)Bacon, Conseil, Mary, Brinchmann, Shepherd,
  Akhlaghi, Weilbacher, Piqueras, Wisotzki, Lagattuta,
  {et~al.}}]{bacon2017muse}
Bacon, R., Conseil, S., Mary, D., {et~al.} 2017, Astronomy \& Astrophysics,
  608, A1

\bibitem[{Bar {et~al.}(2020)Bar, Blum, \& D'Amico}]{Bar:2019ifz}
Bar, N., Blum, K., \& D'Amico, G. 2020, Phys. Rev. D, 101, 123025,
  \dodoi{10.1103/PhysRevD.101.123025}

\bibitem[{Battaglia {et~al.}(2013)Battaglia, Helmi, \&
  Breddels}]{Battaglia:2013wqa}
Battaglia, G., Helmi, A., \& Breddels, M. 2013, New Astron. Rev., 57, 52,
  \dodoi{10.1016/j.newar.2013.05.003}

\bibitem[{{Battaglia} \& {Nipoti}(2022)}]{2022NatAs...6..659B}
{Battaglia}, G., \& {Nipoti}, C. 2022, Nature Astronomy, 6, 659,
  \dodoi{10.1038/s41550-022-01638-7}

\bibitem[{Bessho {et~al.}(2022)Bessho, Ikeda, \& Yin}]{Bessho:2022yyu}
Bessho, T., Ikeda, Y., \& Yin, W. 2022, Phys. Rev. D, 106, 095025,
  \dodoi{10.1103/PhysRevD.106.095025}

\bibitem[{Bonnivard {et~al.}(2015{\natexlab{a}})}]{Bonnivard:2015xpq}
Bonnivard, V., {et~al.} 2015{\natexlab{a}}, Mon. Not. Roy. Astron. Soc., 453,
  849, \dodoi{10.1093/mnras/stv1601}

\bibitem[{Bonnivard {et~al.}(2015{\natexlab{b}})Bonnivard, Combet, Maurin,
  Geringer-Sameth, Koushiappas, Walker, Mateo, Olszewski, \&
  Bailey~III}]{Bonnivard:2015tta}
Bonnivard, V., Combet, C., Maurin, D., {et~al.} 2015{\natexlab{b}}, Astrophys.
  J. Lett., 808, L36, \dodoi{10.1088/2041-8205/808/2/L36}

\bibitem[{Brandt \& Draine(2012)}]{Brandt:2011ka}
Brandt, T.~D., \& Draine, B.~T. 2012, Astrophys. J., 744, 129,
  \dodoi{10.1088/0004-637X/744/2/129}

\bibitem[{Burkert(1995)}]{Burkert:1995yz}
Burkert, A. 1995, Astrophys. J. Lett., 447, L25, \dodoi{10.1086/309560}

\bibitem[{Caputo {et~al.}(2021)Caputo, Vittino, Fornengo, Regis, \&
  Taoso}]{Caputo:2020msf}
Caputo, A., Vittino, A., Fornengo, N., Regis, M., \& Taoso, M. 2021, JCAP, 05,
  046, \dodoi{10.1088/1475-7516/2021/05/046}

\bibitem[{Carenza {et~al.}(2021)Carenza, Lattanzi, Mirizzi, \&
  Forastieri}]{Carenza:2021ebx}
Carenza, P., Lattanzi, M., Mirizzi, A., \& Forastieri, F. 2021, JCAP, 07, 031,
  \dodoi{10.1088/1475-7516/2021/07/031}

\bibitem[{Carenza {et~al.}(2023)Carenza, Lucente, \&
  Vitagliano}]{Carenza:2023qxh}
Carenza, P., Lucente, G., \& Vitagliano, E. 2023, Phys. Rev. D, 107, 083032,
  \dodoi{10.1103/PhysRevD.107.083032}

\bibitem[{Carenza {et~al.}(2020)Carenza, Straniero, D\"obrich, Giannotti,
  Lucente, \& Mirizzi}]{Carenza:2020zil}
Carenza, P., Straniero, O., D\"obrich, B., {et~al.} 2020, Phys. Lett. B, 809,
  135709, \dodoi{10.1016/j.physletb.2020.135709}

\bibitem[{Chang {et~al.}(2018)Chang, Essig, \& McDermott}]{Chang:2018rso}
Chang, J.~H., Essig, R., \& McDermott, S.~D. 2018, JHEP, 09, 051,
  \dodoi{10.1007/JHEP09(2018)051}

\bibitem[{Chang \& Choi(1993)}]{Chang:1993gm}
Chang, S., \& Choi, K. 1993, Phys. Lett. B, 316, 51,
  \dodoi{10.1016/0370-2693(93)90656-3}

\bibitem[{Combet {et~al.}(2012)Combet, Maurin, Nezri, Pointecouteau, Hinton, \&
  White}]{Combet:2012tt}
Combet, C., Maurin, D., Nezri, E., {et~al.} 2012, Phys. Rev. D, 85, 063517,
  \dodoi{10.1103/PhysRevD.85.063517}

\bibitem[{Daido {et~al.}(2017)Daido, Takahashi, \& Yin}]{Daido:2017wwb}
Daido, R., Takahashi, F., \& Yin, W. 2017, JCAP, 05, 044,
  \dodoi{10.1088/1475-7516/2017/05/044}

\bibitem[{Daido {et~al.}(2018)Daido, Takahashi, \& Yin}]{Daido:2017tbr}
---. 2018, JHEP, 02, 104, \dodoi{10.1007/JHEP02(2018)104}

\bibitem[{Dolan {et~al.}(2022)Dolan, Hiskens, \& Volkas}]{Dolan:2022kul}
Dolan, M.~J., Hiskens, F.~J., \& Volkas, R.~R. 2022.
\newblock \doarXiv{2207.03102}

\bibitem[{Drlica-Wagner {et~al.}(2015)}]{DES:2015zwj}
Drlica-Wagner, A., {et~al.} 2015, Astrophys. J., 813, 109,
  \dodoi{10.1088/0004-637X/813/2/109}

\bibitem[{Evans {et~al.}(2016)Evans, Sanders, \&
  Geringer-Sameth}]{Evans:2016xwx}
Evans, N.~W., Sanders, J.~L., \& Geringer-Sameth, A. 2016, Phys. Rev. D, 93,
  103512, \dodoi{10.1103/PhysRevD.93.103512}

\bibitem[{Geringer-Sameth {et~al.}(2015)Geringer-Sameth, Koushiappas, \&
  Walker}]{Geringer-Sameth:2014yza}
Geringer-Sameth, A., Koushiappas, S.~M., \& Walker, M. 2015, Astrophys. J.,
  801, 74, \dodoi{10.1088/0004-637X/801/2/74}

\bibitem[{Giannotti {et~al.}(2016)Giannotti, Irastorza, Redondo, \&
  Ringwald}]{Giannotti:2015kwo}
Giannotti, M., Irastorza, I., Redondo, J., \& Ringwald, A. 2016, JCAP, 05, 057,
  \dodoi{10.1088/1475-7516/2016/05/057}

\bibitem[{Gong {et~al.}(2016)Gong, Cooray, Mitchell-Wynne, Chen, Zemcov, \&
  Smidt}]{Gong:2015hke}
Gong, Y., Cooray, A., Mitchell-Wynne, K., {et~al.} 2016, Astrophys. J., 825,
  104, \dodoi{10.3847/0004-637X/825/2/104}

\bibitem[{Grilli~di Cortona {et~al.}(2016)Grilli~di Cortona, Hardy, Pardo~Vega,
  \& Villadoro}]{GrillidiCortona:2015jxo}
Grilli~di Cortona, G., Hardy, E., Pardo~Vega, J., \& Villadoro, G. 2016, JHEP,
  01, 034, \dodoi{10.1007/JHEP01(2016)034}

\bibitem[{Grin {et~al.}(2007)Grin, Covone, Kneib, Kamionkowski, Blain, \&
  Jullo}]{Grin:2006aw}
Grin, D., Covone, G., Kneib, J.-P., {et~al.} 2007, Phys. Rev. D, 75, 105018,
  \dodoi{10.1103/PhysRevD.75.105018}

\bibitem[{Grin {et~al.}(2008)Grin, Smith, \& Kamionkowski}]{Grin:2007yg}
Grin, D., Smith, T.~L., \& Kamionkowski, M. 2008, Phys. Rev. D, 77, 085020,
  \dodoi{10.1103/PhysRevD.77.085020}

\bibitem[{Hayashi {et~al.}(2020)Hayashi, Chiba, \& Ishiyama}]{Hayashi:2020jze}
Hayashi, K., Chiba, M., \& Ishiyama, T. 2020, Astrophys. J., 904, 45,
  \dodoi{10.3847/1538-4357/abbe0a}

\bibitem[{Hayashi {et~al.}(2018)Hayashi, Fabrizio, \L{}okas, Bono, Monelli,
  Dall'Ora, \& Stetson}]{Hayashi:2018uop}
Hayashi, K., Fabrizio, M., \L{}okas, E.~L., {et~al.} 2018, Mon. Not. Roy.
  Astron. Soc., 481, 250, \dodoi{10.1093/mnras/sty2296}

\bibitem[{Hayashi {et~al.}(2022)Hayashi, Hirai, Chiba, \&
  Ishiyama}]{Hayashi:2022wnw}
Hayashi, K., Hirai, Y., Chiba, M., \& Ishiyama, T. 2022.
\newblock \doarXiv{2206.02821}

\bibitem[{Hayashi {et~al.}(2016)Hayashi, Ichikawa, Matsumoto, Ibe, Ishigaki, \&
  Sugai}]{Hayashi:2016kcy}
Hayashi, K., Ichikawa, K., Matsumoto, S., {et~al.} 2016, Mon. Not. Roy. Astron.
  Soc., 461, 2914, \dodoi{10.1093/mnras/stw1457}

\bibitem[{{Hernquist}(1990)}]{1990ApJ...356..359H}
{Hernquist}, L. 1990, \apj, 356, 359, \dodoi{10.1086/168845}

\bibitem[{Ikeda {et~al.}(2006)Ikeda, Kobayashi, Kondo, Yasui, Motohara, \&
  Minami}]{ikeda2006winered}
Ikeda, Y., Kobayashi, N., Kondo, S., {et~al.} 2006, in Ground-Based and
  Airborne Instrumentation for Astronomy, Vol. 6269, SPIE, 1224--1231

\bibitem[{Ikeda {et~al.}(2016)Ikeda, Kobayashi, Kondo, Otsubo, Hamano,
  Sameshima, Yoshikawa, Fukue, Nakanishi, Kawanishi, {et~al.}}]{ikeda2016high}
Ikeda, Y., Kobayashi, N., Kondo, S., {et~al.} 2016, in Ground-based and
  Airborne Instrumentation for Astronomy VI, Vol. 9908, SPIE, 1800--1813

\bibitem[{Ikeda {et~al.}(2018)Ikeda, Kobayashi, Kondo, Otsubo, Watase, Murai,
  Sakamoto, Hamano, Sameshima, Fukue, {et~al.}}]{ikeda2018very}
Ikeda, Y., Kobayashi, N., Kondo, S., {et~al.} 2018, in Ground-based and
  Airborne Instrumentation for Astronomy VII, Vol. 10702, SPIE, 1751--1762

\bibitem[{{Ikeda} {et~al.}(2022){Ikeda}, {Kondo}, {Otsubo}, {Hamano}, {Yasui},
  {Matsunaga}, {Sameshima}, {Yoshikawa}, {Fukue}, {Nakanishi}, {Kawanishi},
  {Watase}, {Nakaoka}, {Arai}, {Kinoshita}, {Kitano}, {Nakamura}, {Asano},
  {Takenaka}, {Murai}, {Kawakita}, {Minami}, {Izumi}, {Yamamoto}, {Mizumoto},
  {Taniguchi}, \& {Tsujimoto}}]{2022WINERED}
{Ikeda}, Y., {Kondo}, S., {Otsubo}, S., {et~al.} 2022, pasp, 134, 015004,
  \dodoi{10.1088/1538-3873/ac1c5f}

\bibitem[{{Kelsall} {et~al.}(1998){Kelsall}, {Weiland}, {Franz}, {Reach},
  {Arendt}, {Dwek}, {Freudenreich}, {Hauser}, {Moseley}, {Odegard},
  {Silverberg}, \& {Wright}}]{1998IPAC}
{Kelsall}, T., {Weiland}, J.~L., {Franz}, B.~A., {et~al.} 1998, apj, 508, 44,
  \dodoi{10.1086/306380}

\bibitem[{Kobayashi {et~al.}(2000)Kobayashi, Tokunaga, Terada, Goto, Weber,
  Potter, Onaka, Ching, Young, Fletcher, {et~al.}}]{kobayashi2000ircs}
Kobayashi, N., Tokunaga, A.~T., Terada, H., {et~al.} 2000, in Optical and IR
  Telescope Instrumentation and Detectors, Vol. 4008, SPIE, 1056--1066

\bibitem[{Kondo {et~al.}(2015)Kondo, Ikeda, Kobayashi, Yasui, Mito, Fukue,
  Nakanishi, Kawanishi, Nakaoka, Otsubo, {et~al.}}]{kondo2015warm}
Kondo, S., Ikeda, Y., Kobayashi, N., {et~al.} 2015, arXiv preprint
  arXiv:1501.03403

\bibitem[{Koposov {et~al.}(2015)Koposov, Belokurov, Torrealba, \&
  Evans}]{Koposov:2015cua}
Koposov, S.~E., Belokurov, V., Torrealba, G., \& Evans, N.~W. 2015, Astrophys.
  J., 805, 130, \dodoi{10.1088/0004-637X/805/2/130}

\bibitem[{Korochkin {et~al.}(2020)Korochkin, Neronov, \&
  Semikoz}]{Korochkin:2019qpe}
Korochkin, A., Neronov, A., \& Semikoz, D. 2020, JCAP, 03, 064,
  \dodoi{10.1088/1475-7516/2020/03/064}

\bibitem[{Laevens {et~al.}(2015{\natexlab{a}})}]{Laevens:2015kla}
Laevens, B. P.~M., {et~al.} 2015{\natexlab{a}}, Astrophys. J., 813, 44,
  \dodoi{10.1088/0004-637X/813/1/44}

\bibitem[{Laevens {et~al.}(2015{\natexlab{b}})}]{Laevens:2015una}
---. 2015{\natexlab{b}}, Astrophys. J. Lett., 802, L18,
  \dodoi{10.1088/2041-8205/802/2/L18}

\bibitem[{Lutz {et~al.}(2009)}]{PANDA:2009yku}
Lutz, M. F.~M., {et~al.} 2009.
\newblock \doarXiv{0903.3905}

\bibitem[{Martin {et~al.}(2015)}]{Martin:2015xla}
Martin, N.~F., {et~al.} 2015, Astrophys. J. Lett., 804, L5,
  \dodoi{10.1088/2041-8205/804/1/L5}

\bibitem[{Moroi \& Murayama(1998)}]{Moroi:1998qs}
Moroi, T., \& Murayama, H. 1998, Phys. Lett. B, 440, 69,
  \dodoi{10.1016/S0370-2693(98)01091-0}

\bibitem[{Moroi \& Yin(2021{\natexlab{a}})}]{Moroi:2020has}
Moroi, T., \& Yin, W. 2021{\natexlab{a}}, JHEP, 03, 301,
  \dodoi{10.1007/JHEP03(2021)301}

\bibitem[{Moroi \& Yin(2021{\natexlab{b}})}]{Moroi:2020bkq}
---. 2021{\natexlab{b}}, JHEP, 03, 296, \dodoi{10.1007/JHEP03(2021)296}

\bibitem[{{Mu{\~n}oz} {et~al.}(2018){Mu{\~n}oz}, {C{\^o}t{\'e}}, {Santana},
  {Geha}, {Simon}, {Oyarz{\'u}n}, {Stetson}, \& {Djorgovski}}]{Mu_oz_2018a}
{Mu{\~n}oz}, R.~R., {C{\^o}t{\'e}}, P., {Santana}, F.~A., {et~al.} 2018, The
  Astrophysical Journal, 860, 65, \dodoi{10.3847/1538-4357/aac168}

\bibitem[{Mu{\~{n}}oz {et~al.}(2018)Mu{\~{n}}oz, C{\^{o}}t{\'{e}}, Santana,
  Geha, Simon, Oyarz{\'{u}}n, Stetson, \& Djorgovski}]{Mu_oz_2018}
Mu{\~{n}}oz, R.~R., C{\^{o}}t{\'{e}}, P., Santana, F.~A., {et~al.} 2018, The
  Astrophysical Journal, 860, 66, \dodoi{10.3847/1538-4357/aac16b}

\bibitem[{Nakayama \& Yin(2022)}]{Nakayama:2022jza}
Nakayama, K., \& Yin, W. 2022.
\newblock \doarXiv{2205.01079}

\bibitem[{Navarro {et~al.}(1996)Navarro, Frenk, \& White}]{Navarro:1995iw}
Navarro, J.~F., Frenk, C.~S., \& White, S. D.~M. 1996, Astrophys. J., 462, 563,
  \dodoi{10.1086/177173}

\bibitem[{{Navarro} {et~al.}(2010){Navarro}, {Ludlow}, {Springel}, {Wang},
  {Vogelsberger}, {White}, {Jenkins}, {Frenk}, \&
  {Helmi}}]{2010MNRAS.402...21N}
{Navarro}, J.~F., {Ludlow}, A., {Springel}, V., {et~al.} 2010, \mnras, 402, 21,
  \dodoi{10.1111/j.1365-2966.2009.15878.x}

\bibitem[{Peccei \& Quinn(1977{\natexlab{a}})}]{Peccei:1977hh}
Peccei, R.~D., \& Quinn, H.~R. 1977{\natexlab{a}}, Phys. Rev. Lett., 38, 1440,
  \dodoi{10.1103/PhysRevLett.38.1440}

\bibitem[{Peccei \& Quinn(1977{\natexlab{b}})}]{Peccei:1977ur}
---. 1977{\natexlab{b}}, Phys. Rev. D, 16, 1791,
  \dodoi{10.1103/PhysRevD.16.1791}

\bibitem[{Petac {et~al.}(2018)Petac, Ullio, \& Valli}]{Petac:2018gue}
Petac, M., Ullio, P., \& Valli, M. 2018, JCAP, 12, 039,
  \dodoi{10.1088/1475-7516/2018/12/039}

\bibitem[{Raffelt(1986)}]{Raffelt:1985nk}
Raffelt, G.~G. 1986, Phys. Rev. D, 33, 897, \dodoi{10.1103/PhysRevD.33.897}

\bibitem[{Raffelt(1996)}]{Raffelt:1996wa}
---. 1996, {Stars as laboratories for fundamental physics}: {The astrophysics
  of neutrinos, axions, and other weakly interacting particles}

\bibitem[{Raffelt \& Dearborn(1987)}]{Raffelt:1987yu}
Raffelt, G.~G., \& Dearborn, D. S.~P. 1987, Phys. Rev. D, 36, 2211,
  \dodoi{10.1103/PhysRevD.36.2211}

\bibitem[{{Reach} {et~al.}(1997){Reach}, {Franz}, \& {Weiland}}]{1997IPAC}
{Reach}, W.~T., {Franz}, B.~A., \& {Weiland}, J.~L. 1997, icarus, 127, 461,
  \dodoi{10.1006/icar.1997.5704}

\bibitem[{Regis {et~al.}(2021)Regis, Taoso, Vaz, Brinchmann, Zoutendijk,
  Bouch\'e, \& Steinmetz}]{Regis:2020fhw}
Regis, M., Taoso, M., Vaz, D., {et~al.} 2021, Phys. Lett. B, 814, 136075,
  \dodoi{10.1016/j.physletb.2021.136075}

\bibitem[{Sanders {et~al.}(2016)Sanders, Evans, Geringer-Sameth, \&
  Dehnen}]{Sanders:2016eie}
Sanders, J.~L., Evans, N.~W., Geringer-Sameth, A., \& Dehnen, W. 2016, Phys.
  Rev. D, 94, 063521, \dodoi{10.1103/PhysRevD.94.063521}

\bibitem[{Schlegel {et~al.}(1998)Schlegel, Finkbeiner, \&
  Davis}]{Schlegel:1997yv}
Schlegel, D.~J., Finkbeiner, D.~P., \& Davis, M. 1998, Astrophys. J., 500, 525,
  \dodoi{10.1086/305772}

\bibitem[{Straniero {et~al.}(2015)Straniero, Ayala, Giannotti, Mirizzi, \&
  Dominguez}]{Straniero:2015nvc}
Straniero, O., Ayala, A., Giannotti, M., Mirizzi, A., \& Dominguez, I. 2015, in
  {11th Patras Workshop on Axions, WIMPs and WISPs}, 77--81,
  \dodoi{10.3204/DESY-PROC-2015-02/straniero_oscar}

\bibitem[{Takahashi {et~al.}(2021)Takahashi, Yamada, \&
  Yin}]{Takahashi:2020uio}
Takahashi, F., Yamada, M., \& Yin, W. 2021, JHEP, 01, 152,
  \dodoi{10.1007/JHEP01(2021)152}

\bibitem[{Takahashi \& Yin(2021)}]{Takahashi:2021tff}
Takahashi, F., \& Yin, W. 2021, JCAP, 10, 057,
  \dodoi{10.1088/1475-7516/2021/10/057}

\bibitem[{Takahashi \& Yin(2023)}]{Takahashi:2023vhv}
---. 2023.
\newblock \doarXiv{2301.10757}

\bibitem[{Tokunaga {et~al.}(1998)Tokunaga, Kobayashi, Bell, Ching, Hodapp,
  Hora, Neill, Onaka, Rayner, Robertson, {et~al.}}]{tokunaga1998infrared}
Tokunaga, A.~T., Kobayashi, N., Bell, J., {et~al.} 1998, in Infrared
  Astronomical Instrumentation, Vol. 3354, SPIE, 512--524

\bibitem[{Torrealba {et~al.}(2016)Torrealba, Koposov, Belokurov, \&
  Irwin}]{10.1093/mnras/stw733}
Torrealba, G., Koposov, S.~E., Belokurov, V., \& Irwin, M. 2016, Monthly
  Notices of the Royal Astronomical Society, 459, 2370,
  \dodoi{10.1093/mnras/stw733}

\bibitem[{Weinberg(1978)}]{Weinberg:1977ma}
Weinberg, S. 1978, Phys. Rev. Lett., 40, 223,
  \dodoi{10.1103/PhysRevLett.40.223}

\bibitem[{Wilczek(1978)}]{Wilczek:1977pj}
Wilczek, F. 1978, Phys. Rev. Lett., 40, 279, \dodoi{10.1103/PhysRevLett.40.279}

\bibitem[{Yasui {et~al.}(2008)Yasui, Kondo, Ikeda, Minami, Motohara, \&
  Kobayashi}]{yasui2008warm}
Yasui, C., Kondo, S., Ikeda, Y., {et~al.} 2008, in Ground-based and Airborne
  Instrumentation for Astronomy II, Vol. 7014, SPIE, 1095--1106

\bibitem[{Yin(2023)}]{Yin:2023jjj}
Yin, W. 2023.
\newblock \doarXiv{2301.08735}

\bibitem[{{Zhao}(1996)}]{1996MNRAS.278..488Z}
{Zhao}, H. 1996, \mnras, 278, 488, \dodoi{10.1093/mnras/278.2.488}

\bibitem[{Zubko {et~al.}(2004)Zubko, Dwek, \& Arendt}]{Zubko:2003eg}
Zubko, V., Dwek, E., \& Arendt, R.~G. 2004, Astrophys. J. Suppl., 152, 211,
  \dodoi{10.1086/382351}

\end{thebibliography}

\end{document}